\documentclass[useAMS,usenatbib]{mn2e}

\usepackage{graphicx}
\usepackage{siunitx}

\usepackage{microtype}

\title{LOFAR discovery of radio emission in MACS$\,$J0717.5$+$3745 }

\author[A. Bonafede et al.]{A. Bonafede$^{1,2,3},$\thanks{E-mail:
email@address}   M. Br\"uggen$^2$,  D. Rafferty$^2$,  I. Zhuravleva$^{4,5}$,  C. J. Riseley$^6$,
\newauthor R. J. van Weeren$^7$,  J. S. Farnes$^8$, F. Vazza$^{1,2,3}$, F. Savini$^2$, A. Wilber$^2$, A. Botteon$^{1,3}$,
\newauthor G. Brunetti$^1$, R. Cassano$^1$, C. Ferrari$^{10}$, F. de Gasperin$^{2,7}$,  E. Orr\'u$^{11}$,
\newauthor  R.F. Pizzo$^{11}$, H. J. A. R\"ottgering$^{7}$, T.  W. Shimwell$^{7,11}$.\\
$^1$ INAF - Istituto di Radioastronomia, Bologna Via Gobetti 101, I�40129 Bologna, Italy.\\
$^2$ Hamburger Sternwarte, Universit\"at Hamburg, Gojenbergsweg 112, 21029, Hamburg, Germany.\\
$^3$ Dipartimento di Fisica e Astronomia, Universit\'a di Bologna, via P. Gobetti 93/2, 40129, Bologna, Italy.\\
$^4$ Kavli Institute for Particle Astrophysics and Cosmology, Stanford University, 452 Lomita Mall, Stanford, California 94305-4085, USA.\\
$^5$ Department of Physics, Stanford University, 382 Via Pueblo Mall, Stanford, California 94305-4060, USA.\\
$^6$ CSIRO Astronomy \& Space Science, 26 Dick Perry Avenue, Kensington, WA, 6151, Australia.\\
$^7$ Leiden Observatory, Leiden University, PO Box 9513, 2300 RA Leiden, The Netherlands .\\
$^8$ Oxford e-Research Centre (OeRC), Department of Engineering Science, University of Oxford, Oxford, OX1 3QG, UK.\\
$^9$ Department of Astrophysics/IMAPP, Radboud University, PO Box 9010, NL-6500 GL Nijmegen, the Netherlands.\\
$^{10}$ Observatoire de la C\^ote d'Azur, Laboratoire Lagrange UMR 7293. Bd. de l'Observatoire, CS 34229 06304 Nice, France.\\
$^{11}$ ASTRON, the Netherlands Institute for Radio Astronomy, Postbus 2, 7990 AA Dwingeloo, The Netherlands.
}

\begin{document}

\date{Accepted Received...}

\pagerange{\pageref{firstpage}--\pageref{lastpage}} \pubyear{2002}

\maketitle

\label{firstpage}

\begin{abstract}
We present results from LOFAR and GMRT observations of the galaxy cluster MACS$\,$J0717.5$+$3745. The cluster is undergoing a violent merger involving at least four sub-clusters,
and it is known to host a radio halo.  LOFAR observations reveal new  sources of radio emission in the Intra-Cluster Medium: (i) a radio bridge that connects the cluster to a head-tail radio galaxy located along a filament of galaxies falling into the main cluster, (ii) a 1.9 Mpc  radio arc, that is located North West of the main mass component, (iii) radio emission along the X-ray bar, that traces the gas in the X-rays South West of the cluster centre. We use deep GMRT observations  at 608 MHz to constrain the spectral indices of these new radio sources, and of the emission that was already studied in the literature at higher frequency. We find that the spectrum of the radio halo and of the relic at LOFAR frequency follows the same power law as observed at higher frequencies. The radio bridge, the radio arc, and the radio bar all have steep spectra, which can be used to constrain the particle acceleration mechanisms. We argue that the radio bridge could be caused by the re-acceleration of electrons by shock waves that are injected along the filament during the cluster mass assembly.  Despite the  sensitivity reached by our observations, the emission from the radio halo does not trace the emission of the gas revealed by X-ray observations. We argue that this could be due to the difference in the ratio of kinetic over thermal energy of the intra-cluster gas, suggested by X-ray observations.

\end{abstract}


\section{Introduction}
The intra-cluster medium (ICM) of galaxy clusters is filled with a weakly magnetised plasma that can contain 
relativistic electrons emitting synchrotron radiation over Mpc scale.
The emission is classified as radio halos and radio relics depending on its morphological properties: radio halos are found at cluster centres and are mostly connected with major mergers  (e.g. \citealt{Buote01,Cuciti15}, but see also \citealt{Bonafede14b} and \citealt{Sommer17} for some outliers), while radio relics usually have an arc-like morphology and are found at cluster peripheries \citep[e.g][]{vanWeeren10}. \par
Mergers between galaxy clusters can dissipate up to $10^{64}$ ergs  of energy in the ICM, and current theoretical models predict that a fraction of this energy may be channeled into the (re)acceleration of cosmic-ray electrons (CRe) and magnetic field amplification. Turbulence injected by mergers could re-accelerate a population of low energy CRe and produce radio halos, while shock waves that propagate in the ICM during mergers could amplify the magnetic field and (re)accelerate CRe producing radio relics. We refer the reader to the reviews by \citet{ Brueggen11,Feretti12,BJ14} for more details.\par
 MACSJ0717.5+3745 (hereinafter MACSJ0717) is a very complex system. It is located at  redshift $z=0.546$, and since its discovery, it has been the subject of several observational campaigns \citep[e.g.][]{2003MNRAS.339..913E,Ma08,Bonafede09b,vanWeeren09,Medezinski13,Sayers13,Umetsu14,Limousin16,vanWeeren16_macsA,Adam17}. The main properties of the cluster are listed in Table \ref{tab:macs}.\par
 X-ray and optical observations show a complex merger involving  at least four sub-clusters \citep[e.g][]{Limousin16}. The ICM temperature shows strong gradients, with the eastern part significantly hotter than the western part. The hottest region in the SE reaches $\sim$ 20 keV, while recent {\it Chandra } data suggest the presence of a cold front in the  N-NE region\citep{vanWeeren17} . The X-ray emission reveals a V-shaped structure, associated with the main mass component. To the North West of this component, a bullet-like structure is associated with a second sub-cluster. On the SE, a bar-shaped structure coincides with two more sub-clusters (see Fig. \ref{fig:Xradio}, \ref{fig:gimp}).\par
In the South-East of the cluster, a 19 Mpc long filament of galaxies was found by \citet{Ebeling04} and confirmed by \citet{Jauzac12}. Recently, deep {\it Chandra} observations have detected the part of the filament that is close to the cluster, and found a galaxy group of $\sim$ 10$^{13} \, M_{\odot}$ embedded in the filament  (Ogrean et al. 2016). A  head-tail radiogalaxy (hereinafter HT radiogalaxy) at $z=0.5399$ is found along the filament,  between the cluster and the X-ray detected group  \citep{Ebeling14}.  In Fig. \ref{fig:Xradio}, the X-ray emission is shown and the different components of the system are labelled.\\
 We refer the reader to \citet{Ebeling14, Limousin16, Medezinski13, vanWeeren17}, for a detailed analysis of the cluster X-ray emission and dynamical state. \par
 The cluster hosts a powerful radio halo \citep{Bonafede09b,vanWeeren09}, that is asymmetric and whose largest linear size is more than 1.4 Mpc.
 Within  the halo, the cluster also hosts a bright polarised filament or radio relic, aptly named the ``chair-shaped" filament by \citet{PandeyPommier13}. For simplicity, we will refer to this structure as a relic. The relic is polarised at the $\sim$ 17\% level at 4.9 GHz, while the halo also shows polarisation at the 2-7 \% at 1.4 GHz \citep{Bonafede09b}.
 Such polarised emission -- which is common for relics -- is unusual for halos, and may be related to the peculiar dynamical state of the system.\par
The emission from the radio halo roughly follows the bar and V- shape structures detected in the X-rays, while no emission was found in the western part of the cluster \citep{Bonafede09b,vanWeeren09}. Deep {\it Jansky} Very Large Array (VLA) observations have recently found  several radio filaments on scales 100-300 kpc,  departing from the halo towards the NE and the NW, and at least a few of these are located in the cluster outskirts \citep{vanWeeren17}.\par
Thanks to its high radio power, this is one of the few radio halos that can be imaged by existing interferometers at frequencies higher than 1.4 GHz. Hence, it is a primary target to study the spectral properties of the radio emission over a large frequency range.
\citet{vanWeeren17} found an average spectral index\footnote{Throughout this paper we define the spectral index $\alpha$ as $ S(\nu) \propto \nu^{\alpha}$.} of $\sim$ -1.3  to -1.4 by fitting a straight power law through flux measurements at 1.5, 3.0, and 5.5 GHz . This is in agreement with previous results by \citet{Bonafede09b}, and \citet{PandeyPommier13}, obtained with shallower and lower frequencies observations, respectively.\par
In this paper, we present new low frequency observations of the cluster obtained with the LOw Frequency ARray (LOFAR, \citealt{LOFAR}).
 Our aim is to constrain the spectral properties of the diffuse emission to gain insights on the (re)acceleration processes in this complex system, and to search for additional emission in the western part of the cluster where, despite the presence of hot gas and dynamical activity, no radio emission has been detected.\par
 The remainder of the paper is as follows:
 in Sec. \ref{sec:observations}, we describe the radio observations and the main steps of the data reduction. In Sec. \ref{sec:results}, we analyse the
 results of the LOFAR and Giant Meterwave Radio Telescope (GMRT) observations, and discuss the spectral properties of the system.
 A combined radio and X-ray analysis is reported in Sec. \ref{sec:radioX}, and we conclude in Sec. \ref{sec:discussion}.
   Throughout the paper we use a $\Lambda$CDM cosmological model with $H_0=$ 69.6 km s$^{-1}$ Mpc$^{-1}$, $\Omega_m=$0.286, $\Omega_{\Lambda}=$0.714.
 At the cluster redshift the angular to linear scale is  6.459 kpc/$''$.
 
 \begin{table}
\label{tab:macs}
 \centering
   \caption{MACSJ0717.5+3745}
  \begin{tabular}{l c c}
  \hline
Name &   MACSJ0717.5+3745 & ref. \\
RA  [J2000] & 07h17m30.9s  & \\
DEC[J2000] &37d45$'$30$''$ &\\
$z$	 &	0.546 & E01 \\
$\rm{M_{500}}$  &(11.5 $\pm$ 0.5) $\times$ 10$\rm{^{14}}  M_{\odot}$ &PUC \\ 
$\rm{M_{vir}}$  & (3.5 $\pm$ 0.6 $\times$ 10$\rm{^{15}}  M_{\odot}$  & U14 \\ 
$L_X [0.1-2.4 \rm{keV}]$  & $(2.74 \pm 0.03) \times 10^{45}$  erg/s & E07 \\ 
\hline
\hline

\multicolumn{3}{l}{\scriptsize E01:  \citet{Ebeling01}, PUC: \citet{PUC}}\\
\multicolumn{3}{l}{\scriptsize U14:  \citet{Umetsu14}, E07: \citet{Ebeling07}.}\\
\end{tabular}
 \end{table}

\section{Observations and data reduction}

\label{sec:observations}
\subsection{LOFAR}
\label{LOFAR}
The cluster was observed on 2013 March 19 for a total of 5 hours, using the LOFAR High Band Antenna (HBA) stations in the HBA$\_$DUAL$\_$INNER mode. A total of 61 antennas were present (13 Remote stations, and 24 Core stations, each split into two). Observations covered the frequency range 111 $-$ 182 MHz using 366 sub-bands.
3C286 was observed in the frequency range 115 $-$ 176 MHz using 310 sub-bands and was used as calibrator. Both the cluster and the calibrator data were taken with a sampling time of 2s. Each sub-band of 0.195 MHz bandwidth was recorded with 64 channels.  Data were initially flagged by the observatory using AOFlagger \citep{aoflagger} and then averaged down to 4 channels per sub-band and 5 and 4 s sampling for the cluster and calibrator, respectively. \par
Data were calibrated using the facet calibration approach \citep{facetcal}. We refer the reader to \citet{facetcal} for a detailed description, and outline here only the main steps. \par
Calibrator data were further averaged to 10s and 2 channels per sub-band. The stations CS013HBA and CS032HBA were flagged and the
data were calibrated against a source model, following the \citet{ScaifeHeald12} flux density scale. XX and YY gains were determined, together with the rotation angle to account for possible differential Faraday rotation, which was found to be negligible. 
Clock offsets for each  station were derived from these solutions, fitting for Clock delays and differential TEC (Total Electron Content). 
 Amplitude gains, clock offsets, and instrumental XX-YY phase-offsets were applied to the target data to set the initial flux scale, and to correct for instrumental effects, respectively. 
 
 Then, the target data were calibrated in phase against a {\it Global Sky Model}, that is derived from several radio surveys (specifically, the VLA Low-frequency Sky Survey, VLSS,  \citealt{VLSSr},  the WEsterbork Northern Sky Survey, WENSS, \citealt{WENSS}, and the Northern VLA Sky Survey, NVSS \citealt{NVSS}). \par 
A first set of images at intermediate resolution ($\sim$ 30$''$) were created using WSClean \citep{wsclean,wsclean-2017}, grouping the data in chunks of $\sim$ 2  MHz each.  Images were corrected for the station beam at the phase centre.
The model components were subtracted from the UVdata and new images at lower resolution  ($\sim 2'$) were created, including sources up to 20 degrees from the target centre. The model component list was updated with the clean components found in the low resolution images.  After this step, we are left with a list of model components for the field, and an almost empty UV dataset, as required by the {\it Factor} pipeline\footnote{https://github.com/lofar-astron/factor.}, that performs the facet calibration. \par
The LOFAR HBA field of view has been divided into 50 facets. Each facet is set to contain a source brighter than 0.1 Jy and
smaller than 2 arcmin, that is used as facet calibrator. UVdata from baselines shorter than 80 $\lambda$ have not been used during self-calibration and
deconvolution. Our aim is to derive the direction-dependent gains in the direction of the target. To do this, we first need to minimise the artefacts from bright sources around the target. The contamination due to artefacts from nearby sources is modest, and indeed the cluster itself is the second brightest source in the field after {\it  B3 0704+384}.
Hence, we first derived direction-dependent gains in the direction of  {\it B3 0704+384}. These gains have been applied to the facet and a new model has been derived and subtracted from the visibilities. 
 Direction-dependent gains were then derived for the target facet using the cluster as calibrator. 
We also checked  that deriving direction-dependent gains for the facets around the target before processing the target facet did not lead to a better calibration for the target, because these facets are affected by residual artefacts and calibration errors from the cluster facet.
 Direction-dependent gains are derived through several self-calibration cycles, using a multi-resolution algorithm. In this procedure, we initially image the data at 20$''$ resolution and progressively increase the resolution to $\sim 5''$. 
Facets are processed as follows:
data are phase-shifted towards the center of the facet calibrator and further averaged in frequency to speed up the calibration process. The model components of the facet calibrator are added back to the visibilities and several cycles of Stokes I phase and TEC self-calibration are performed on a 10 s time scale. Finally, some rounds of complex gain self-calibration are performed.\par
\begin{figure*}
\includegraphics[width=8.5cm]{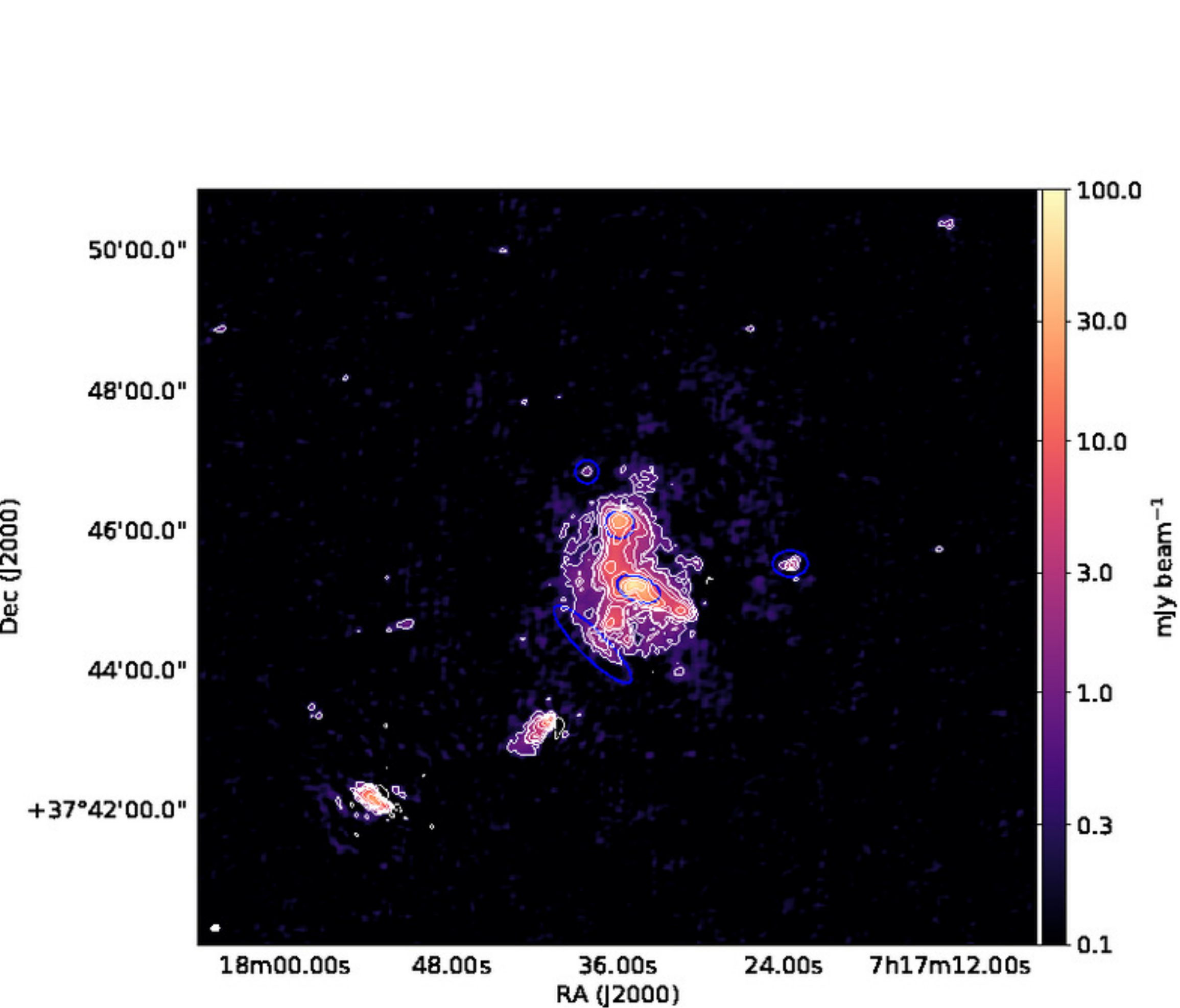}
\includegraphics[width=8.5cm]{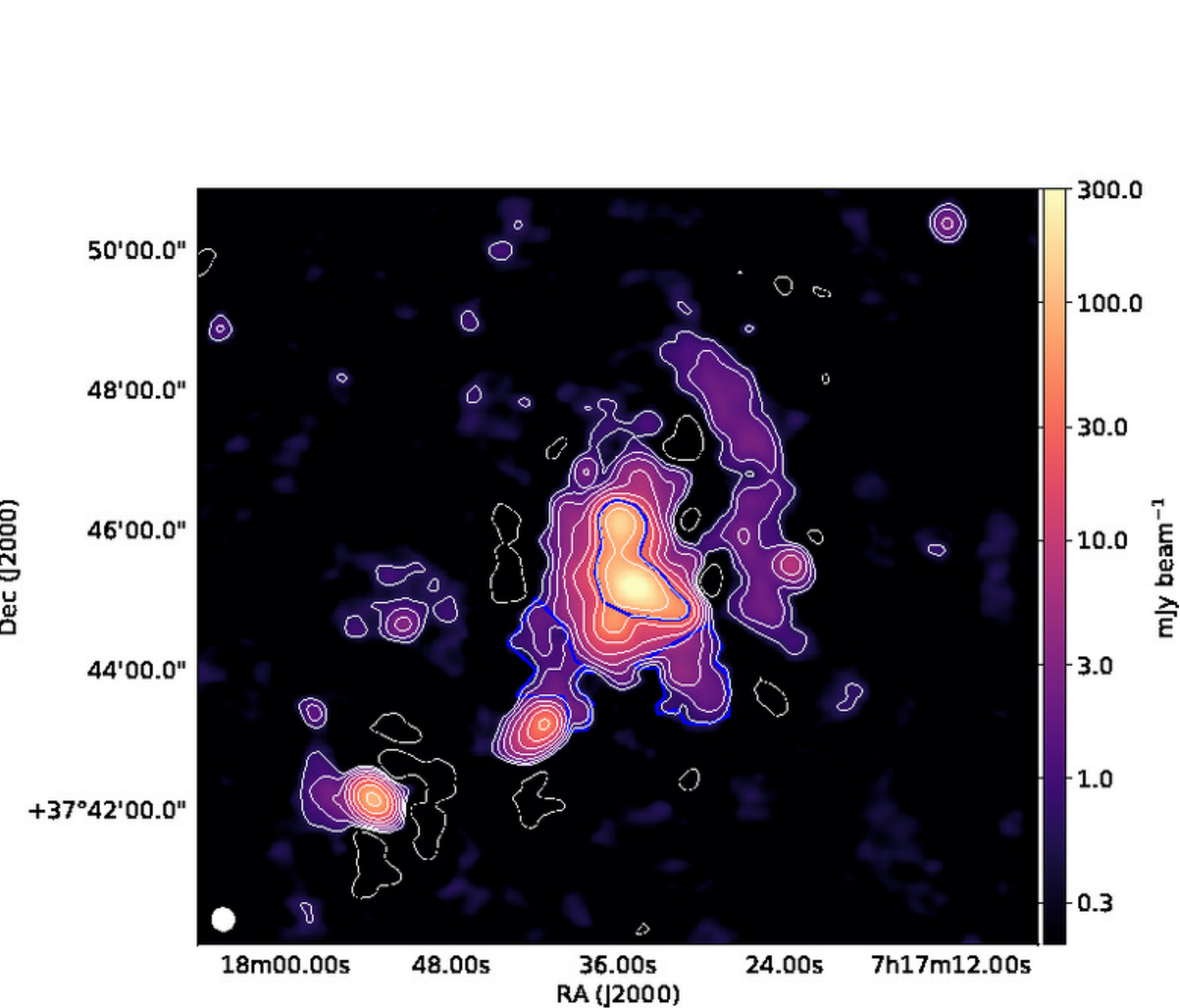}
\caption{Left panel: LOFAR image at 147 MHz at high resolution ($5.8'' \times 4.6''$). The rms noise is 0.13 mJy/beam. Blue circles and ellipses indicate the sources embedded in the diffuse emission, that were masked.
 Right panel: LOFAR image at 147 MHz at low resolution ($19'' \times 18''$). The noise is 0.16 mJy/beam. Blue polygons mark the regions used to separate the different components of the radio emission (bridge, bar, arc) 
In both panels contours start at 4$\sigma$ and are spaced by a factor 2. Contour at -4$\sigma$ are dashed. The restoring beams are shown in the bottom-left corner of the two panels.  }
\label{fig:LOFAR}
\end{figure*}

After calibration, data were imaged using the Briggs weighting scheme \citep{Briggs} setting the parameter $robust=-0.25$ to suppress the sidelobes. Gaussian tapers were applied to decrease the weight of long baselines and better image the extended emission.  Images were done with WSClean using the multi-scale and  multi-frequency deconvolution mode implemented in the code. Imaging parameters are listed in Table \ref{tab:images}, and the images are shown in Fig. \ref{fig:LOFAR}.
\par 
Flux densities have been checked against the TGSS \citep{TGSS_adr}, and we found that the LOFAR flux densities are consistent within the calibration errors. For consistency with previous works, we adopt a conservative 15\% flux density uncertainty \citep[e.g][]{vanWeeren17,Wilber17,Savini17,Shimwell16}.

\begin{table*}

 \centering
   \caption{Images.}
  \begin{tabular}{c c c c c c }
  \hline
Image  		& Weighting scheme & UV-Taper & Restoring beam & rms noise  & Fig of merit\\
		    		&	  &        &   & mJy/beam  & 		\\				
				&&&&\\	
LOFAR HBA HR	&  Briggs, robust=-1            &	-         & 5.8$'' \times$4.6$''$         &  0.13   & \ref{fig:LOFAR}\\
LOFAR HBA LR	& Briggs, robust= -0.25 	 & 10	$''$	&   19$'' \times$18$''$          &  0.16 &  \ref{fig:LOFAR} \\
GMRT 608 HR		& Briggs, robust =0.25       & $- $    &  6.3$ '' \times$5.6$''$    &  0.03       &  \ref{fig:GMRT}\\
GMRT 608 LR		& Briggs, robust= -0.25     & 20$''$   &  18.3$'' \times$17.4$''$    &   0.10 &  \ref{fig:GMRT}\\\\
&&&&&\\
\multicolumn{6}{c}{For spectral index image $-$ UV-range$> 500 \lambda$}\\
&&&&\\
Image 		& Weighting scheme & UV-Taper & Restoring beam & rms noise  & Fig of merit\\
		    		&	  &        &   & mJy/beam  & 		\\
LOFAR HBA	&Uniform  & 10 & 10$''\times $10$''$    &   0.25 &\ref{fig:spix},  top panel \\
GMRT 608   	&Uniform  & 10 &10$''\times $10$''$    &     0.07 &\ref{fig:spix}, top panel  \\
LOFAR HBA	&Uniform  & 30 &30$''\times $30$''$    &     0.50 &\ref{fig:spix}, bottom panel\\
GMRT 608        &Uniform  & 30& 30$''\times $30$''$   &     0.17  &\ref{fig:spix}, bottom panel \\
\hline
\hline

\end{tabular}
\label{tab:images}
\end{table*}

\begin{figure*}
\includegraphics[width=15cm]{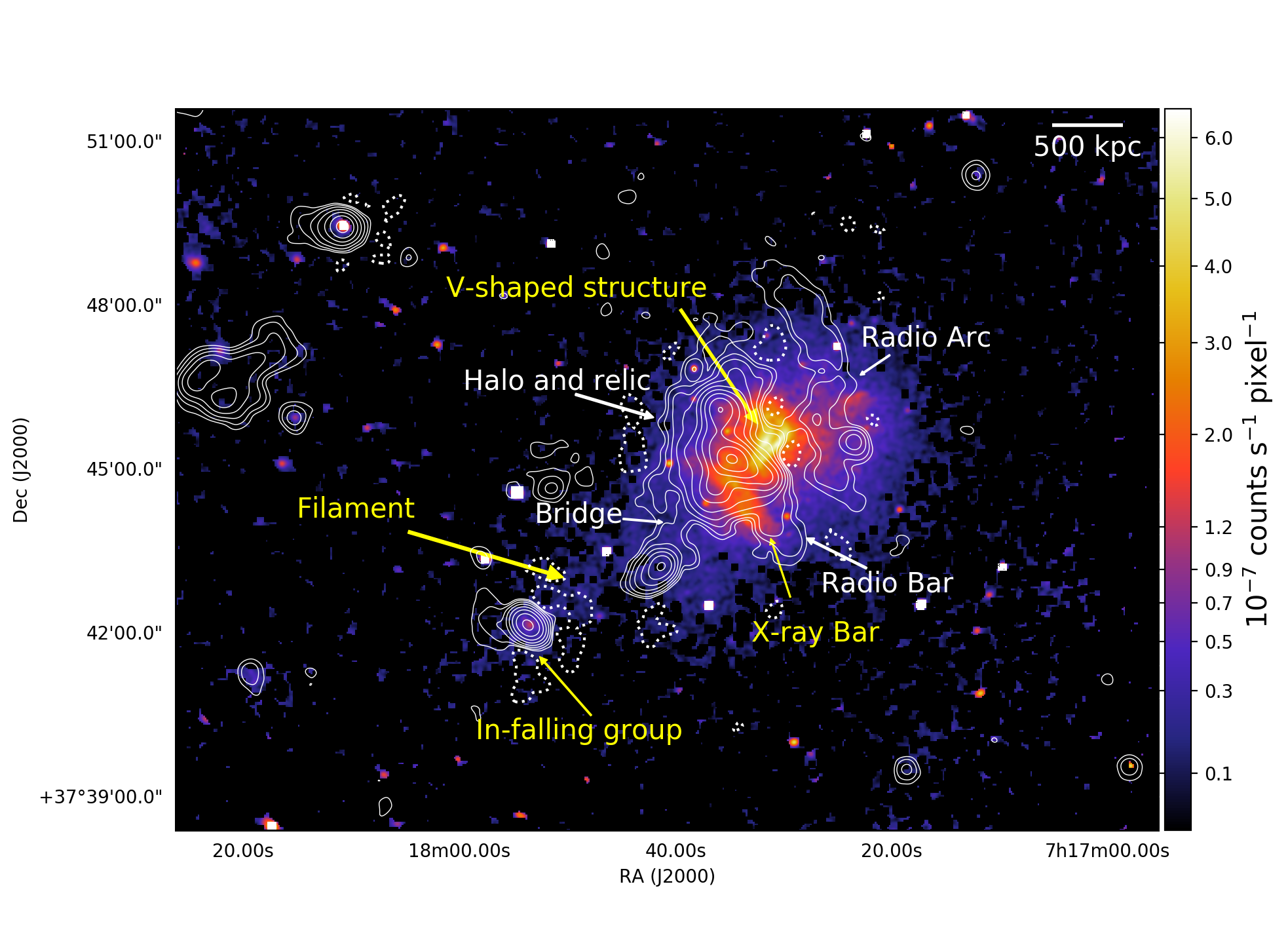}
\caption{Colours: X-ray emission from {\it Chandra} in the band 0.2 $-$ 5 keV. Contours: radio emission from LOFAR at 147 MHz. The beam is 19"$\times$18". The rms noise ($\sigma$) is 0.16 mJy/beam. Contours  start at 4$\sigma$  and are spaced by a factor 2. The contour at $-4\sigma$ is dashed. The main components of the cluster emission in the X-rays and radio are labelled in yellow and white, respectively.}
\label{fig:Xradio}
\end{figure*}

\begin{figure*}
\centering
\includegraphics[width=13cm]{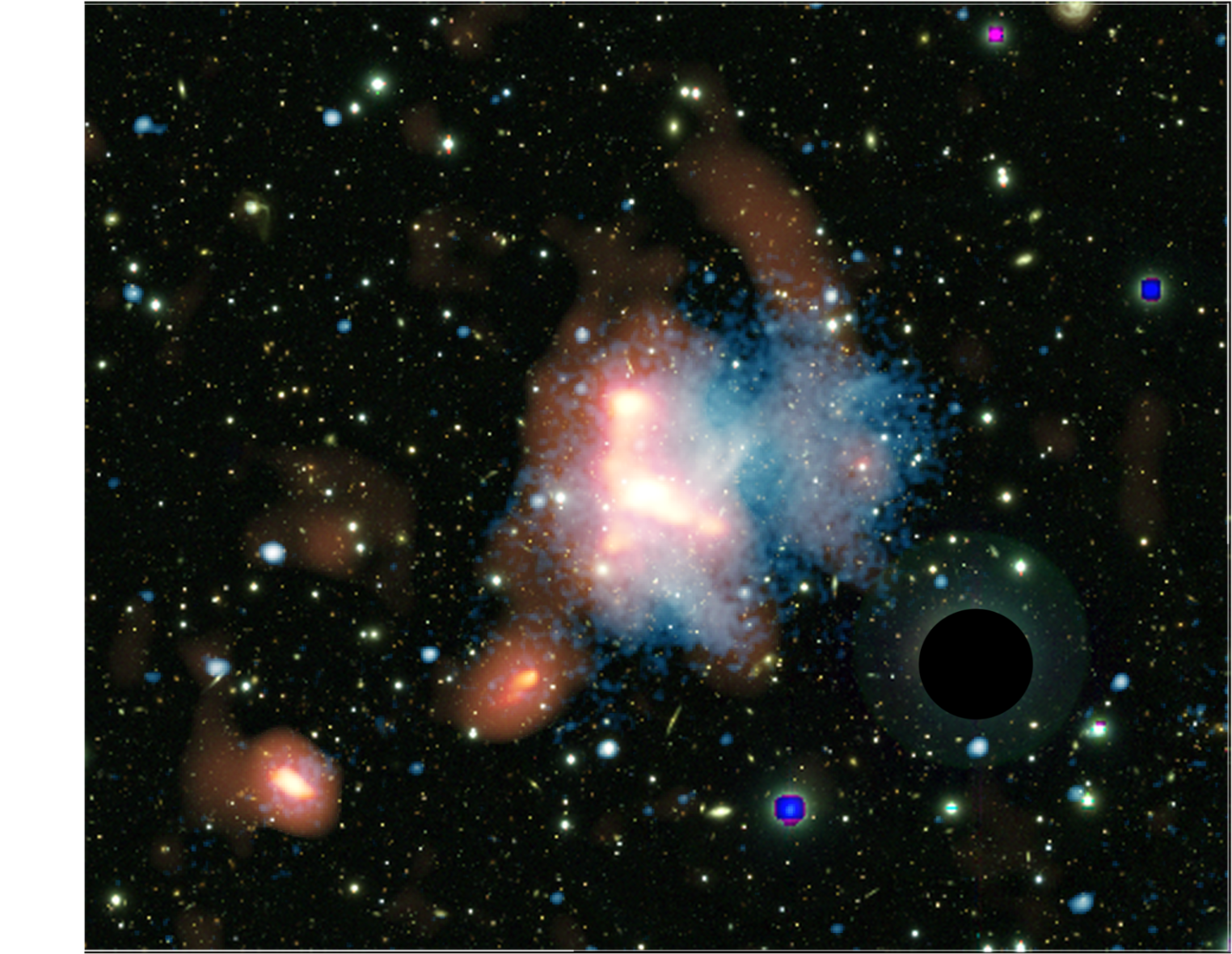}
\caption{Subaru {\it b,v,z} composite image overlaid onto X-ray emission from Chandra (cyan) in the band 0.2 $-$ 5 keV, and radio emission from LOFAR at 147 MHz (orange).
{ The image has illustrative purposes only. A black circle is superposed to a bright star to mask it. Quantitative values for the X-ray and radio emission are shown in Fig. \ref{fig:Xradio} and \ref{fig:LOFAR}, respectively. }}
\label{fig:gimp}
\end{figure*}

\begin{figure*}
\includegraphics[width=8.5cm]{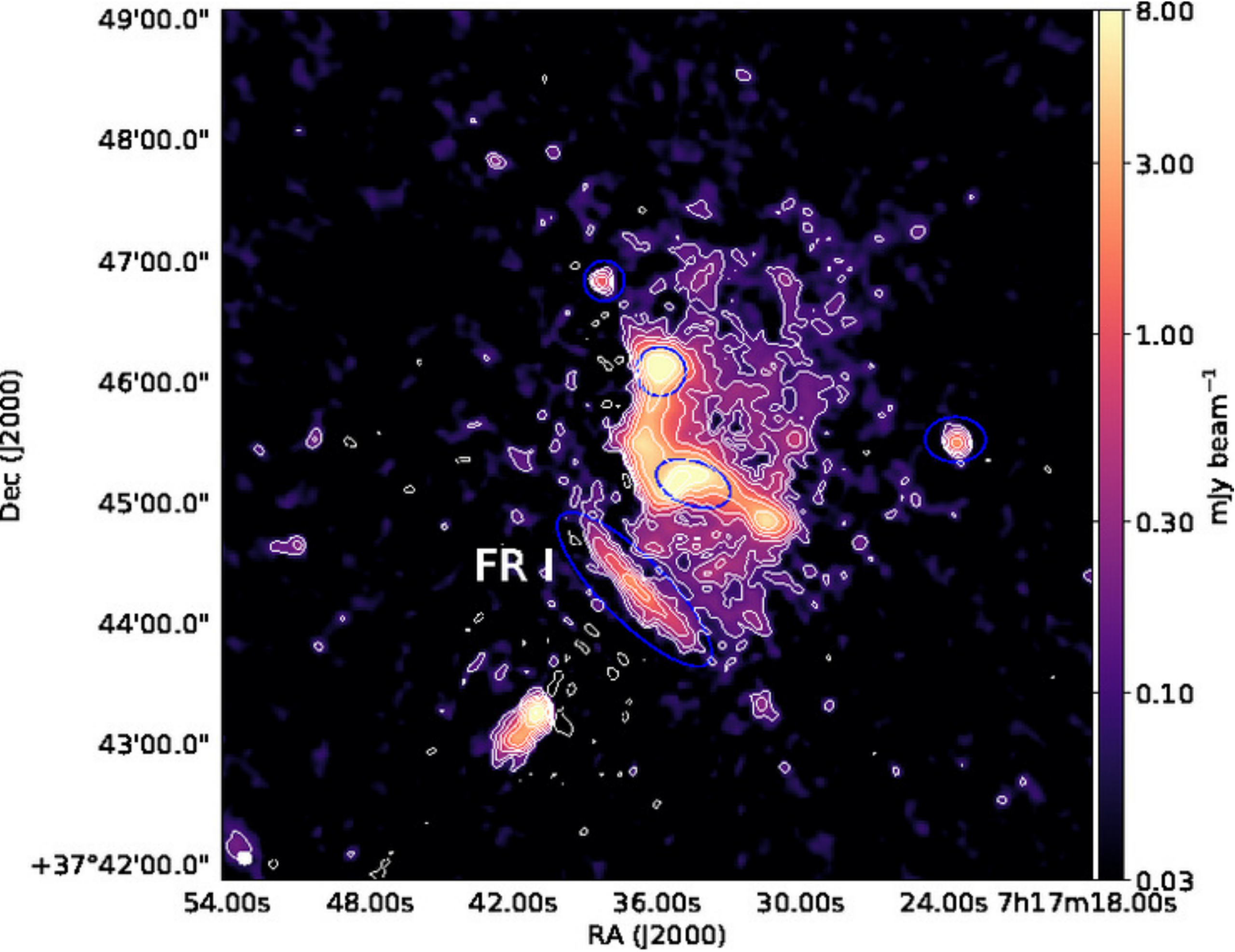}
\includegraphics[width=8.5cm]{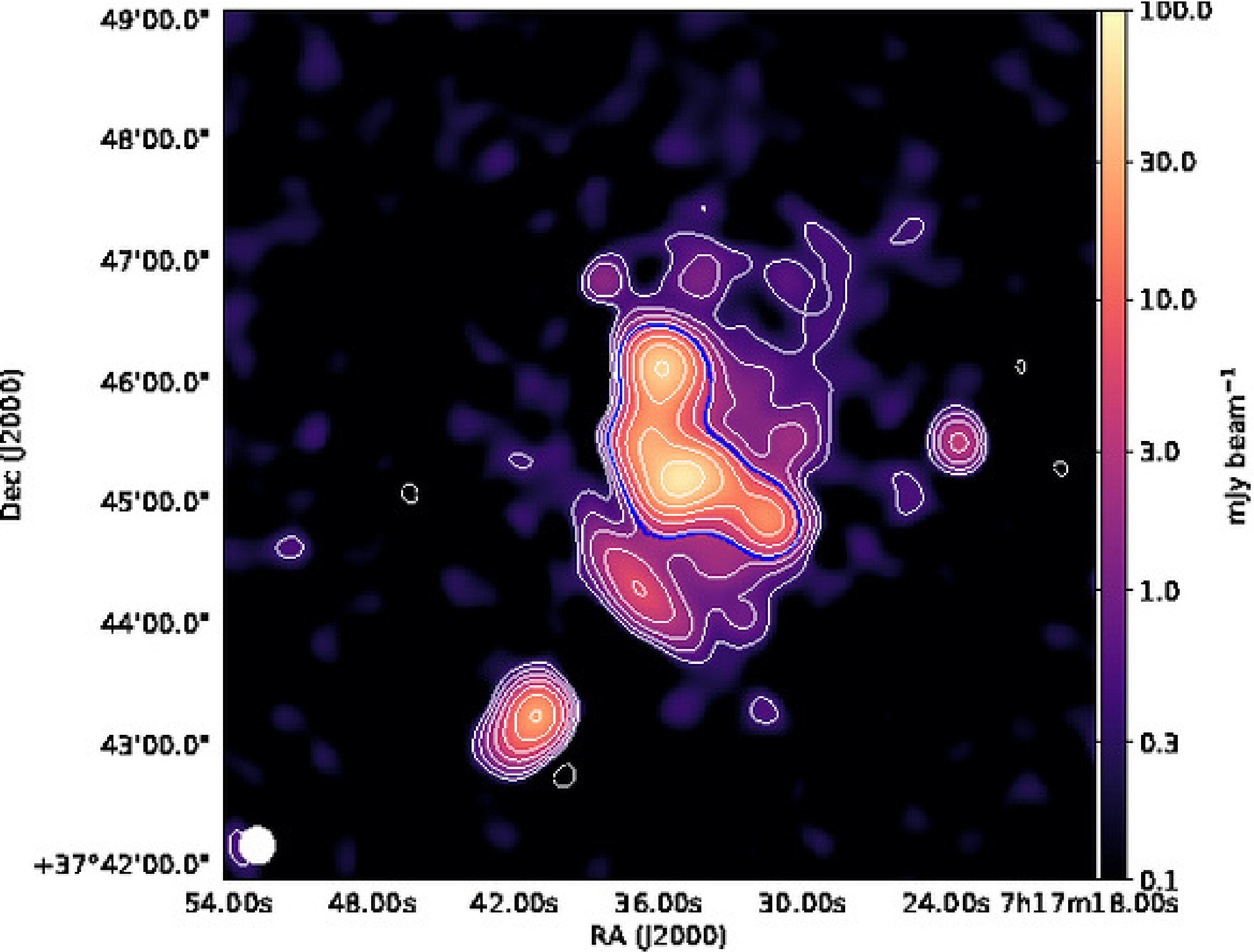}

\caption{Left panel: GMRT image at 608 MHz at high resolution ($6.3'' \times 5.6''$). The rms noise is 30$\mu$Jy/beam. Right panel: GMRT image at 608 MHz at low resolution ($18.3'' \times 17.4''$). The noise is 0.1 mJy/beam. Blue circles and ellipses indicate the sources embedded in the diffuse emission, that have been masked.
In both panels, contours start at 4$\sigma$ and are spaced by a factor 2. Contour at -4$\sigma$ are dashed. The restoring beams are shown in the bottom-left corner of the two panels.}
\label{fig:GMRT}
\end{figure*}

\subsection{GMRT observations}
A 12 h long observation at 608 MHz was performed with the GMRT on June, 5th 2011. \par
Data have been recorded in spectral mode, using 256 channels having a width of 130 kHz each, for a total bandwidth of 33 MHz. The integration time was set to 8 s.
{ We processed the observation using the Source Peeling and Athmospheric Modeling (SPAM) tool} \citep{SPAM} to take into account direction-dependent effects. The main steps are outlined below, and we refer the reader to \citet{SPAM} for further details.\par
The sources 3C147 and 3C286 were observed for 25 min at the beginning and at the end of the observing block, respectively, and used
to correct for the bandpass and to set the absolute flux scale, following the \citet{ScaifeHeald12} flux scale.
{ Strong radio-frequency interference (RFI) } were removed from the data using statistical outlier flagging tools, and much of the remaining low-level RFI was modelled and subtracted from the data using Obit \citep{obit}. After RFI removal, data were averaged down to 24 channels, to speed up the following steps and, at the same time, avoid significant bandwidth smearing during imaging. To correct for the phase gains of the target field, we started from a global sky model (see Sec. \ref{LOFAR}).  SPAM permits to correct for ionospheric effects, and remove direction-dependent gain errors, reaching thermal-noise limited images.
Within SPAM, imaging is done with AIPS using the wide-field imaging technique to compensate for the non-complanarity of the array. 
The presence of strong sources in the field of view enables one to derive directional-dependent gains for each of them (similar to the peeling technique)
and to use these gains to fit a phase-screen over the entire field of view. After ionospheric corrections, sources outside the inner $8'$ were subtracted to facilitate 
the imaging steps.\par
Data have been imaged with CASA \citep{casa} using different weighting schemes and Gaussian UV-tapers to achieve different resolutions.
The final images have been corrected for the GMRT primary beam response, and are shown in Fig. \ref{fig:GMRT}. Imaging parameters are listed in Table \ref{tab:images}. 
We assume a 10\% error on the absolute flux scale.

\section{Results}
\label{sec:results}
{ Both LOFAR and GMRT observations detect new emission that was not detected by previous, shallower radio observations.} Because of the different sensitivities
 of the two instruments towards large-scale emission, we first analyse the observations separately, and then perform a spectral index study.
 
\subsection{Radio emission at 147 MHz} 

The main result of the LOFAR observations is the discovery of additional emission
W of the halo (radio arc), around the X-ray bar, and SE of the halo in the direction of the accreting sub-group along the intergalactic filament (bridge).
The halo emission is more extended than previously found by VLA and GMRT observations \citep{Bonafede09b,PandeyPommier13,vanWeeren17} and it extends beyond the relic in the E and SE directions (Fig. \ref{fig:Xradio}). 
In Fig. \ref{fig:Xradio}, the radio emission at 147 MHz is shown in contours, and the new features are labelled.\par

A foreground  FRI radio galaxy (z=0.1546) has been identified to the SE of the cluster. 
Its lobe are prominent in the VLA and GMRT observations  (see Fig. \ref{fig:GMRT}), but are almost undetected in the LOFAR image because of the combined effect of their spectral index ($\alpha^{\rm 6.5 GHz}_{\rm 1 GHz} \sim $ 0.6, \citealt{vanWeeren17} ) and low surface brightness ($\sim$ 0.2 mJy/beam at 5 GHz, \citealt{vanWeeren17}).\par

The radio arc has a total flux density of 49 $\pm$ 7 mJy. Assuming  $\alpha=-1.3$ for the  $k-$correction, this flux density 
corresponds to a power
$\rm{P_{147 MHz}} = (7 \pm 1) \times 10^{25}$ W/Hz.
The largest angular size of the emission is $\sim$ 4.9$'$, corresponding to $\sim$1.9 Mpc. 
The arc is located at a projected distance of $\sim 1.2'$ (460 kpc) from the main mass component, in-between the W sub-cluster and the V-shaped emission visible in the X-rays.
The radio emission from the radio arc does not follow the X-ray emission from the gas in the same region of the cluster.
The structure could be seen in projection onto the cluster centre, and could be associated to a
merger or accretion shock. In this case, it could be classified as a relic.  However, given the complex structure of the whole radio emission and having no information about projection effects, any conclusion would be speculative. \par
SE of the X-ray bar, hints of a new radio bridge are found, connecting the radio halo to the HT radiogalaxy located along the optical filament $\sim 3'$ from the X-ray centre\footnote{As no evidence for radio emission from the bridge was found in the VLA or in the GMRT observations \citep{Bonafede09b,vanWeeren17,PandeyPommier13}, and as the background noise of the images are not uniform, even a detection at $\sim$ 10 $\sigma$ could be partially affected by calibration artefacts. Hence, deeper observations with other instruments would be required to confirm with higher confidence the properties of the radio bridge and of the radio arc.}.
Bridges connecting radio halos to HT radiogalaxies have been found in few other clusters already (e.g. the Coma cluster \citealt{Giovannini93}) suggesting that the fossil electrons from the tail are (re)accelerated by phenomena connected with the merger. However, this case is somewhat different. In fact, the lobes of the HT radiogalaxy are pointing in the opposite direction with respect to the bridge. The properties of the radio bridge -- in connection with the gas properties of the filament -- are further analysed in Sec. \ref{sec:bridge}.\\
\citet{vanWeeren17} have found that the radio emission SE of the main mass concentration roughly follows the X-ray bar. LOFAR observations reveal further emission covering the entire bar and extending beyond it.

\subsection{Radio emission at 608 MHz}
The GMRT observation allows us to reach a sensitivity that is a factor  $\sim$ 2 deeper with respect to the data published 
so far at this frequency  \citep{vanWeeren09}.
In Fig. \ref{fig:GMRT}, the emission at 608 MHz is shown at two different resolutions, obtained with the imaging parameters listed in Table \ref{tab:images}.
The GMRT image at low resolution shows diffuse emission that was previously undetected at this frequency:
the halo appears more extended in the NW direction, and towards S. No emission is detected corresponding to the radio arc and to the bridge visible in the LOFAR images. 
However, we note that the halo extension toward NW, which is also partially detected in the VLA image at 1.4 GHz \citep{Bonafede09b} is not detected by LOFAR. \par
\smallskip
In Table \ref{tab:data}, we list the flux densities and sizes of the radio components in the cluster both at LOFAR and GMRT frequencies. 
These are derived from the low resolution images (see Tab. \ref{tab:images} for details), above the 4$\sigma$ contour, and masking the discrete sources embedded in the diffuse emission.
Since the boundaries of the relic and of the bar cannot be easily separated from the
halo component,  we also list the properties of the total radio emission, which includes the halo, the relic, and the radio bar.
In Fig. \ref{fig:LOFAR} and \ref{fig:GMRT}, we show the regions used to compute the flux densities listed in Table  \ref{tab:data}.


 \begin{table*}\label{tab:fluxes}
 \centering
   \caption{MACSJ0717.5+3745: details on radio components}
  \begin{tabular}{l c c c c c}
  \hline
Radio component & $S_{\rm 147 MHz } $ &  $S_{\rm 608 MHz } $ &  Size at 147 MHz$^*$    & Size  at 608 MHz$^{*}$ &  $ \alpha^{\rm 147 \, MHz }_{\rm 608 \, MHz}$ \\
			& Jy					& Jy				& arcsec $-$ kpc			& arcsec $-$ kpc \\
Total	 		  & 	0.9 $\pm$	0.1				& 0.21  $\pm$ 0.03 &                                   &\\
Filament/Relic	  &	0.50  $\pm$	0.08			       &   0.18  $\pm$  0.03   &   125 $-$ 800         &   125  $-$ 807 & $-1.1 \pm 0.1$\\
Halo  		  &     0.37 $\pm$    0.06                	       &    0.034 $\pm$   0.005 &       160 $-$ 1030  &	160 $-$ 1030 & $-1.4 \pm 0.1$	\\
Bridge		 &$\rm{(1.3  \pm  0.5) \times 10^{-2}}$   &          $-$      &	65 $-$ 400               &  $-$  & $< -1.4$\\
Bar		         &	0.020 $\pm$	0.003    &        $\rm{(1.7  \pm  0.2) \times 10^{-3}}$   &       70 $-$ 450   	  &  10 $-$  65 & $< -1.6$\\ 
Arc			&     0.049  $\pm$ 0.007 &			-	&  	300 $-$ 1900 	  &  & $< -1.3$ \\
\hline
\hline
\multicolumn{6}{l}{\scriptsize  Col1: Name of the radio source; Col 2 and 3 : Flux density from the LOFAR LR and GMRT LR image;}\\
\multicolumn{6}{l}{\scriptsize    Col 4 and Col 5: Maximum angular and linear size of the source in LOFAR and GMRT. $^*$For the Halo, the parameter $D_H$, defined as $\sqrt{D_{max}\times D_{min}}$}\\
 \multicolumn{6}{l}{\scriptsize  is given, where $D_{max}$ and $D_{min}$ refer to the maximum and minimum scale, respectively. All quantities are projected. }\\
  \multicolumn{6}{l}{\scriptsize  Col. 6: spectral index between 608 MHz and 147 MHz. Note that the spectral index refers to the regions as specified in the text, and  }\\
  \multicolumn{6}{l}{\scriptsize   not to the entire size of the radio component listed in Col 3 and 4.}

\end{tabular}
\label{tab:data}
 \end{table*}

\begin{figure*}
\includegraphics[width=18cm]{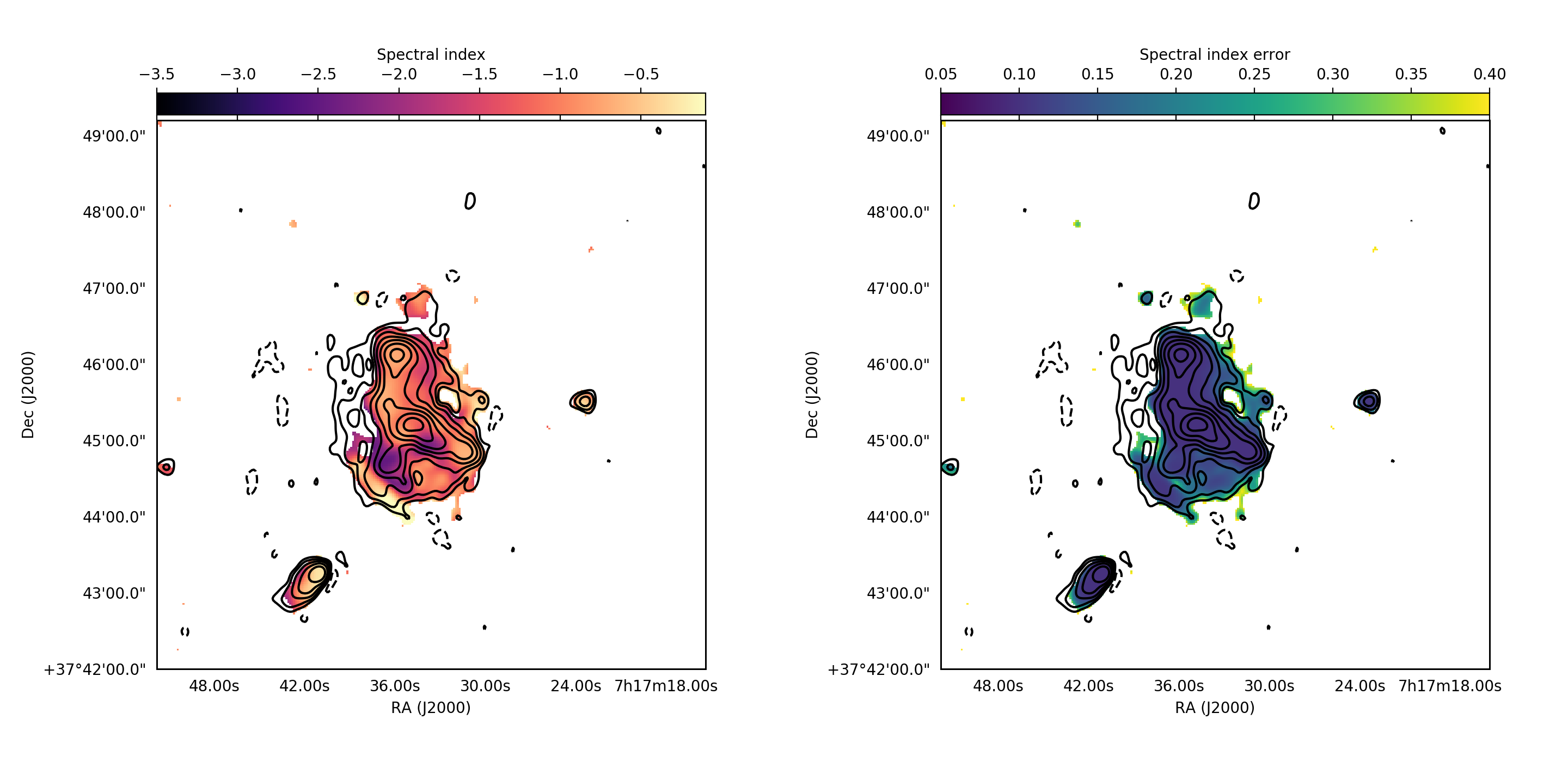}
\includegraphics[width=18cm]{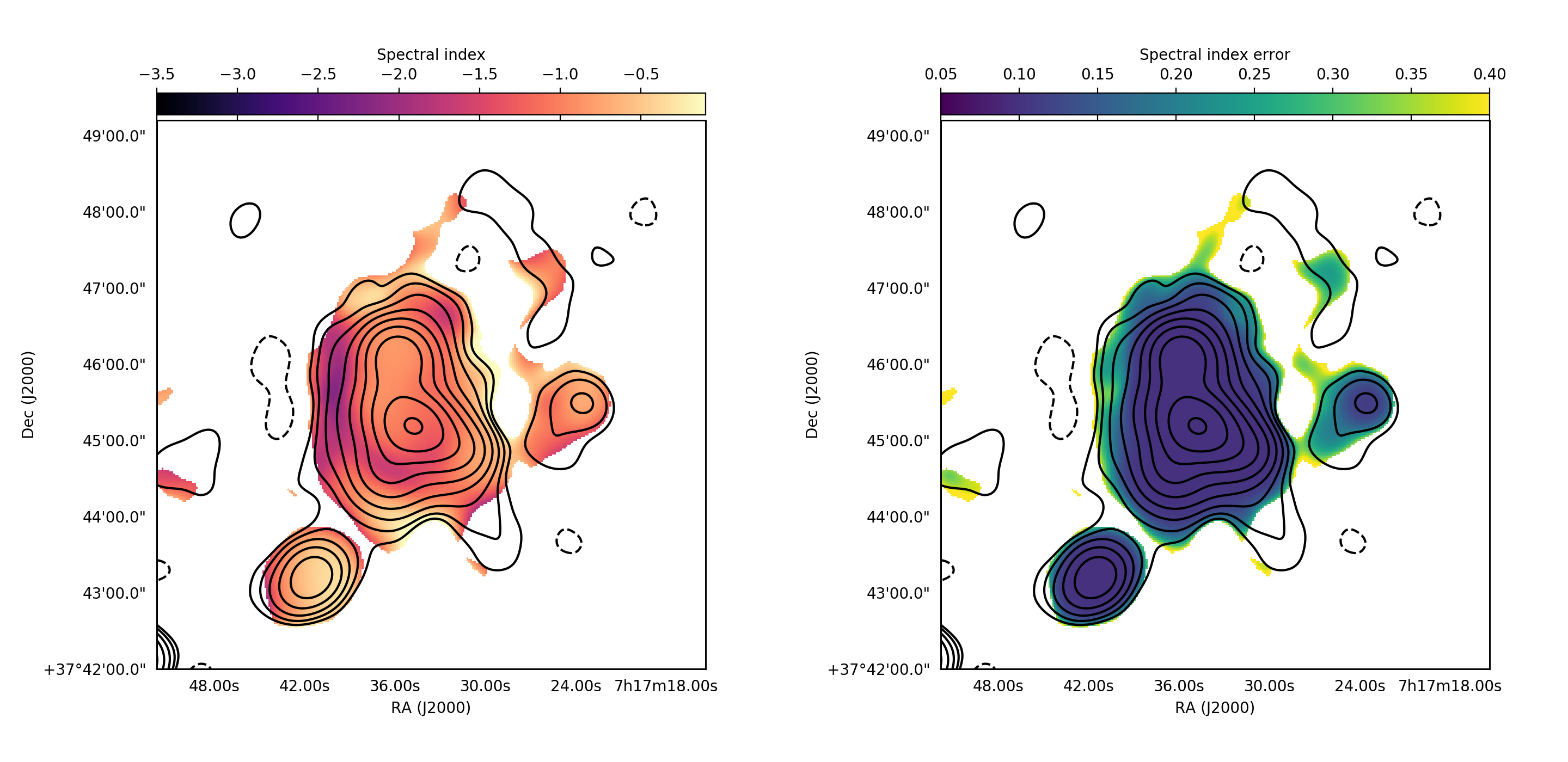}
\caption{Top: LOFAR-GMRT spectral index image (left) and associated errors (right) at the resolution of 10 arcsec. 
Bottom:  LOFAR-GMRT spectral index image (left) and associated errors (right) at the resolution of 30 arcsec.
Contours display the LOFAR stokes I image, starting at $4\sigma$
and increasing of a factor 2 each. The -4$\sigma$ contours are plotted with dotted lines.}
\label{fig:spix}
\end{figure*}
\subsection{Spectral properties of the radio emission}

Using LOFAR and GMRT observations, we have produced spectral index maps of the cluster radio emission. LOFAR and GMRT observations have been imaged 
using the same UV-range, uniform weighting scheme, and a Gaussian taper as listed in Table \ref{tab:images}. The minimum baseline has been chosen to have a 
dense sampling of the GMRT data, and it is particularly critical here, given the different frequency and baseline lengths of the two interferometers. \par
We have used two tapering functions with different FWHM of 10 and 30 arcsec in order to analyse the spectral index variations, and to 
constrain the spectral index of the diffuse emission, respectively. 
To compute the spectral index image, LOFAR and GMRT images have been convolved with a Gaussian beam to achieve the exact same resolution.
Both images have been blanked at 2$\sigma$ and a spectral index map has been  
computed. The spectral index images are shown in Fig. \ref{fig:spix}.  The errors on the spectral index have been computed according to
\begin{equation}
\alpha_{err}= \frac{1}{\ln (\nu_1/\nu_2)}\sqrt{\left( \frac{\Delta S_1}{S_1}\right )^2 + \left( \frac{\Delta S_2}{S_2}\right )^2  },
\end{equation}
\par
where $\Delta S_i$ takes into account both the flux density errors ($\delta S_i \times S_i$) and the image noises ($\sigma_i$). \par
The high resolution spectral index image shows that the cluster diffuse emission is steep, in agreement with previous works  by \citet{Bonafede09b,vanWeeren09,vanWeeren17}.  We detect a steep spectrum region in-between the relic and the foreground FRI radiogalaxy with a mean value $\langle \alpha^{\rm 147 \, MHz }_{\rm 608 \, MHz} \rangle=-2.2 \pm 0.2$. This region could be contaminated by the past emission of a steep-spectrum source detected at higher frequency by \citet{vanWeeren17}.
The Narrow Angle Tail (NAT) radiogalaxy at the cluster centre shows a steepening of the spectral index along the tail  from $\alpha^{\rm 147 \, MHz }_{\rm 608 \, MHz} \sim -0.7$  in the core down to $\alpha^{\rm 147 \, MHz }_{\rm 608 \, MHz} \sim -2.3$, consistent with the behaviour at higher frequencies reported by \citet{vanWeeren17}. Further out along the relic the spectral index becomes flatter ($\alpha^{\rm 147 \, MHz }_{\rm 608 \, MHz} \sim -1.3$ to $ -1$), possibly indicating that the aged electrons from the radio galaxy have been re-energised by a shock  \citep{vanWeeren17}. \par
To compute an average spectral index of the radio halo and of the radio relic, we have blanked the sources embedded in the the diffuse emission. The average spectral index of the halo is 
$\langle \alpha^{\rm 147 \, MHz }_{\rm 608 \, MHz} \rangle =-1.4 \pm 0.1$. We note that this value is the emission detected both at 608 and 147 MHz and it is not representative of the whole emission detected by LOFAR or GMRT.
The average spectral index of the relic is $\langle \alpha^{\rm 147 \, MHz }_{\rm 608 \, MHz} \rangle =-1.1 \pm 0.1$. These values are consistent with those reported in the literature by \citet{Bonafede09b,vanWeeren09,vanWeeren17} within the errors, and indicate that the spectrum does not change significantly at LOFAR frequencies. \par

The new features detected by LOFAR (i.e.  the radio arc, the radio bridge, and the radio bar) are not visible in the GMRT image. This is likely due to a combination of their steep spectrum, weak surface brightness, and large angular extent which is filtered out by the GMRT. 
Imaging the data using the same UV-range and  restoring beam allows us to investigate this.
 In Fig. \ref{fig:spix}, the spectral index at low resolution is shown, together with the LOFAR contours. From this image, we can conclude that: (i) the spectrum of the emission E of the relic is steep ($\langle \alpha^{\rm 147 \, MHz }_{\rm 608 \, MHz} \rangle = -1.7 \pm 0.2 $). (ii) Most of  the radio arc is not detected in the GMRT image, because of both its low surface brightness and large-scale size. Indeed, only the brightest patch of emission is seen by LOFAR once baselines shorter than 500 $\lambda$  are excluded from imaging. In this region, we can put a limit $\langle \alpha^{\rm 147 \, MHz }_{\rm 608 \, MHz} \rangle< -1.3 $.
(iii) The radio bridge and the radio bar are not detected in the GMRT image because of their steep spectrum. We can put a limit on the spectral index in these regions  $\langle \alpha^{\rm 147 \, MHz }_{\rm 608 \, MHz} \rangle < -1.6$ and $\langle \alpha^{\rm 147 \, MHz }_{\rm 608 \, MHz} \rangle  < -1.4$ in the radio bar and radio bridge, respectively. All the limits to the spectra computed above consider the mean LOFAR surface brightness and 2$\sigma$ noise of the GMRT image. \par
The emission NW of the radio halo, detected by GMRT observations, is not visible in the LOFAR image. As the LOFAR image is dominated by the bright halo and radio arc, we cannot exclude that the emission is not visible because of deconvolution artefacts. Alternatively, this emission would need a spectral index $\alpha \geq -1.3 $ to fall below the LOFAR sensitivity.

\section{Radio  and X-ray emission}
\label{sec:radioX}
Using deep {\it Chandra} observations of the cluster (Ogrean et al., 2016; \citealt{vanWeeren17}), we can derive constraints on the particle acceleration mechanisms that produce the
radio emission in the cluster centre and outskirts. In this section, we perform a joint radio and X-ray analysis of the radio bridge and  the radio halo.

\subsection{The radio bridge}
\label{sec:bridge}
The detection of radio emission along the filament connecting the HT radiogalaxy with the main cluster allows us to constrain the non-thermal properties
at the outskirts of clusters. The radio emission is detected in a region between the main cluster and a sub-group which has a temperature of $\sim$ 3 keV and a X-ray luminosity of $\sim 10^{43} \rm{ergs/s}$ in the band 0.1 $-$ 2.4 keV. The group is located at 2 Mpc SE from the main cluster and it is likely at its first infall towards the cluster (Ogrean et al. 2016).
{ The portion of the filament between the group and the cluster is overdense by a factor 100 -- 150 with respect to the critical density of  the Universe at the redshift of the cluster.
This part of the filament has a temperature of  $1.6^{+0.5}_{-0.3}$ keV and a density of $\sim 10^{-4} \, \rm{cm^{-3}}$ (Ogrean et al. 2016)}. \par
{ Being within $r_{100}  - r_{150}$}, the radio emission in the bridge is probing a region that is gravitationally bound to the main cluster, where the magnetic field has likely  been compressed and amplified, erasing all signatures from a primordial seed \citep[e.g.][and ref. therein]{Dolag02,Miniati15}. The detection of radio emission indicates that relativistic electrons are present 
in this region. The central galaxy of the group is radio loud, but being at its first infall onto the main cluster, it is unlikely that it contributes to the radio emission in the bridge. 
 \par
During the accretion of matter onto the main cluster, shock waves are injected in the ICM, that can heat the gas, and accelerate particles through Fermi-I type mechanisms, like Diffusive Shock Acceleration \citep[DSA, e.g.][]{Drury83}.

Using the constraints on the gas temperature and density derived from X-ray observations (Ogrean et al. 2016), we investigate here whether and under which conditions
the radio emission in the bridge can be produced by shock (re)acceleration through DSA.

We start by investigating a simple ``single zone" model for the shock, in which we characterise the entire bridge region with a single value for the gas density, gas temperature, and magnetic field. 
We numerically solve the time-dependent evolution of the energy distribution of relativistic electrons under the following conditions: (i) shocks accelerate electrons through DSA, and the resulting energy distribution of the particles is a power-law in energy ($\frac{dN}{d E} \propto E^{-\delta}$),  with $\delta$ that depends on the injection Mach number, $\mathcal{M_{\rm inj}}$, according to  $\delta=2( \mathcal{M}^2+1)/(\mathcal{M}^2-1)$, e.g.  \citealt{1999ApJ...520..529S}). (ii) Particles undergo energetic losses due to synchrotron and Inverse Compton, as well as collisional losses. (iii)  Electrons might be re-accelerated by a second shock, shortly before the epoch of our radio observation.
In the linear acceleration regime, the particle post-shock spectrum  after re-acceleration will be \citep{Markevitch05,ka12}
\begin{equation}
\frac{dN}{d \gamma} = (\delta +2) \gamma^{-\delta} \int_{\gamma_{min} }^{\gamma} {\frac{dN_a}{d\gamma} \gamma^{\delta-1} d\gamma},
\end{equation}
where $\gamma$ is the Lorentz gamma factor of electrons, $\gamma_{min}$ is the minimum $\gamma$ factor of the particle injected by the first shock after their ageing,
and $\frac{dN_a}{d\gamma}$ is the spectrum of the aged electrons. The scenarios where particles are re-accelerated by a second shock are labelled with $+ \, re$ in Tab. \ref{tab:sim} and Fig. \ref{sim_pred}.\par
The time-dependent diffusion-loss equation of cosmic ray electrons \citep[e.g.][]{1962SvA.....6..317K,1999ApJ...520..529S} is solved with the \citet{1970JCoPh...6....1C} finite difference scheme, using 5$\times$10$^4$ energy bins of $\Delta \gamma=10$ in the $2 \leq \gamma \leq 5 \cdot 10^4$ energy range, and a fixed timestep of $10$ Myr. \par
As the size of the filament in the X-rays is bigger than the size of the radio bridge,  we can not
disentangle whether the average values of density and temperature in the filament are  pre or post shock values. Hence, we have tested both scenarios (labelled with $pre-$ and $post-$, respectively in Tab. \ref{tab:sim} and Fig. \ref{sim_pred}).
We have applied jump conditions as a function of the assumed Mach number  ($\mathcal{M}_{\rm inj}$) to recover the pre-shock values of $n_e$ and $T$. 
In the re-acceleration scenarios, the Mach number of the second shock ($\mathcal{M}_{\rm re}$) is used to compute the pre-shock density and temperature values.
The kinetic energy flux , $\Phi$, through the shock surface is proportional to $\Phi \propto \rho_{\rm pre} \mathcal{M}^3 c_{\rm s,pre}^3$, where $\rho_{\rm pre}$ is the pre-shock gas mass density and $c_{\rm s,pre}$ is the pre-shock sound speed  \citep[e.g.][]{va15radio}.\par
We have investigated different combinations of magnetic fields, Mach numbers and times of injection and re-acceleration needed to reproduce the flux density of the radio bridge at 147 MHz and the radio spectral index ($\alpha^{\rm 147 \, MHz }_{\rm 608 \, MHz} \leq -1.4$). 
In Table  \ref{tab:sim}, we list the main parameters of our modell:  the post-shock gas density and temperature values ($n_{\rm post}$ and $T_{\rm post}$), the  Mach number of the first shock ($\mathcal{M}_{\rm inj}$) at the epoch $t_{\rm inj}$, and  the efficiency of the first shock acceleration $\xi_{\rm e,inj}$. 
The resulting energy in relativistic electrons is $E_{\rm e,inj}$. \\ For the re-acceleration models, we also list the Mach number of the re-accelerating shock ($\mathcal{M}_{\rm re}$) active at the epoch $t_{\rm re}$. 
We list in the Table also the magnetic field of the radio bridge, $B$, the flux density at 147 MHz, $S_{147 \, \rm{MHz}}$, the radio spectral index $\alpha^{\rm 147 \, MHz }_{\rm 608 \, MHz}$, and the predicted flux density at 50 MHz, $S_{50 \,\rm{MHz}}$. The radio emission is obtained by numerically integrating the synchrotron emission from the final distribution of accelerated particles \citep[e.g.][]{1965ARA&A...3..297G}.
Figure \ref{sim_pred} shows the expected flux densities for the above models as a function of the observing frequency. \par

{ Although all models are tailored to reproduce the observed flux density and spectral index limit, some of them can be ruled out: 
single injection scenarios (i.e. $pre$ and $post$ models), as well as $pre + re$ scenario require either a large injection efficiency, which is troublesome for DSA \citep[e.g.][]{va15relics}, and/or  high values of temperature and density that are not compatible with the observational constraints.

The $post + re$ scenarios is the only one, among those investigated here, that could reproduce the radio bridge flux density with reasonable values of the model parameters. 
In this model, the second shock has a Mach number $\mathcal{M}_{\rm re}=3$, that would produce radio emission on a scale $L_{rad}$ which is much smaller than the projected size of the bridge (see e.g. eq. 15 in \citealt{ka12}). Hence, one should assume that the shock propagates with a very small angle with respect to the line of sight, perpendicular to the filament main axis. 
Such ``transversal" shocks are observed in cosmological simulations in filaments that connect interacting clusters \citep{va15radio}.
In this case, the observed spectrum will be the superposition of different populations of electrons (re)accelerated at slightly different times as the shock propagates  through the radio bridge \citep[][]{1999ApJ...520..529S}.\par
We have resimulated this scenario by considering the emission from the superposition of the different populations of electrons (scenario $post+re2$ in Table \ref{tab:sim}). While most parameters are unchanged (see the last row of Table \ref{tab:sim}),  one needs to assume a larger energy of reaccelerated electrons to compensate for the cooling losses of the layers that have been accelerated first. 
The predicted spectrum in this case becomes flatter at higher frequencies (Fig.\ref{sim_pred}), leaving to future observations the possibility to better investigate this scenario.\par

\par 
\begin{figure}

\includegraphics[width=0.45\textwidth]{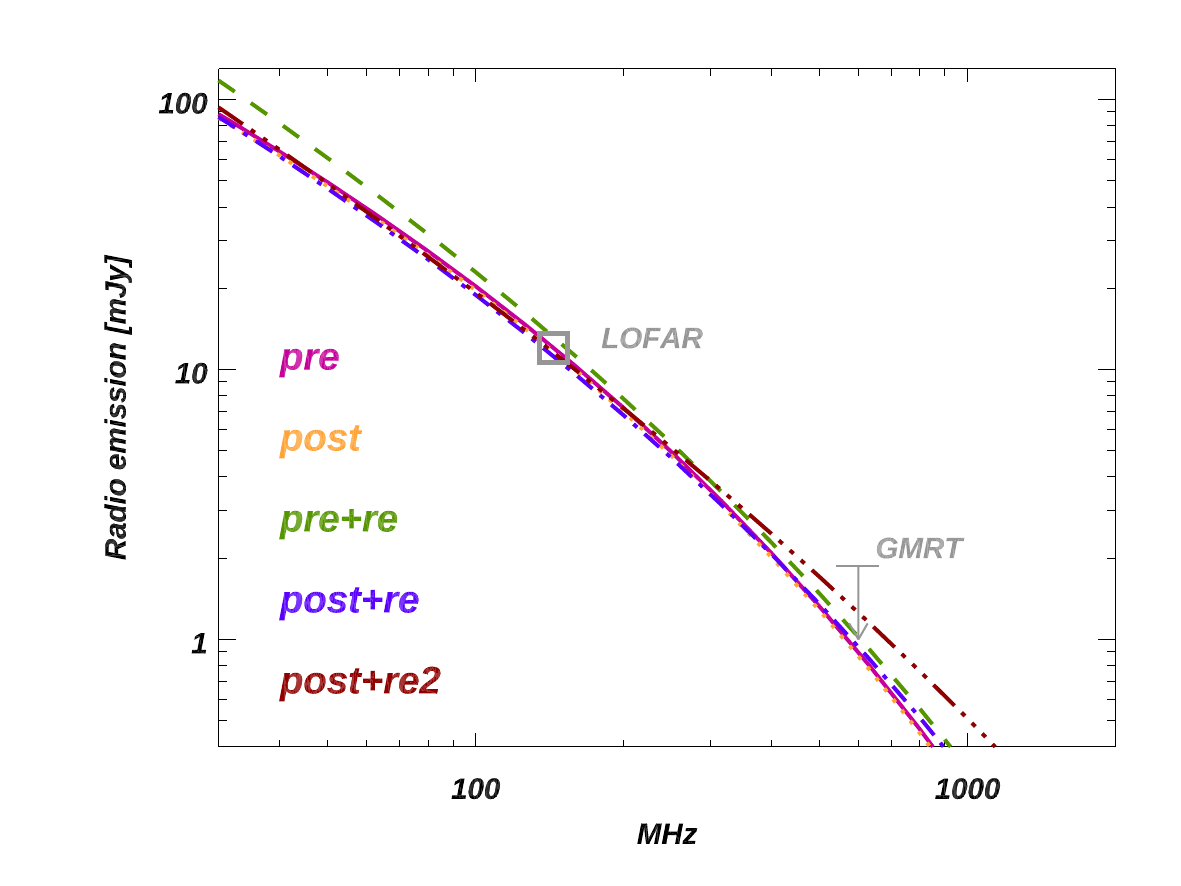}
\includegraphics[width=0.45\textwidth]{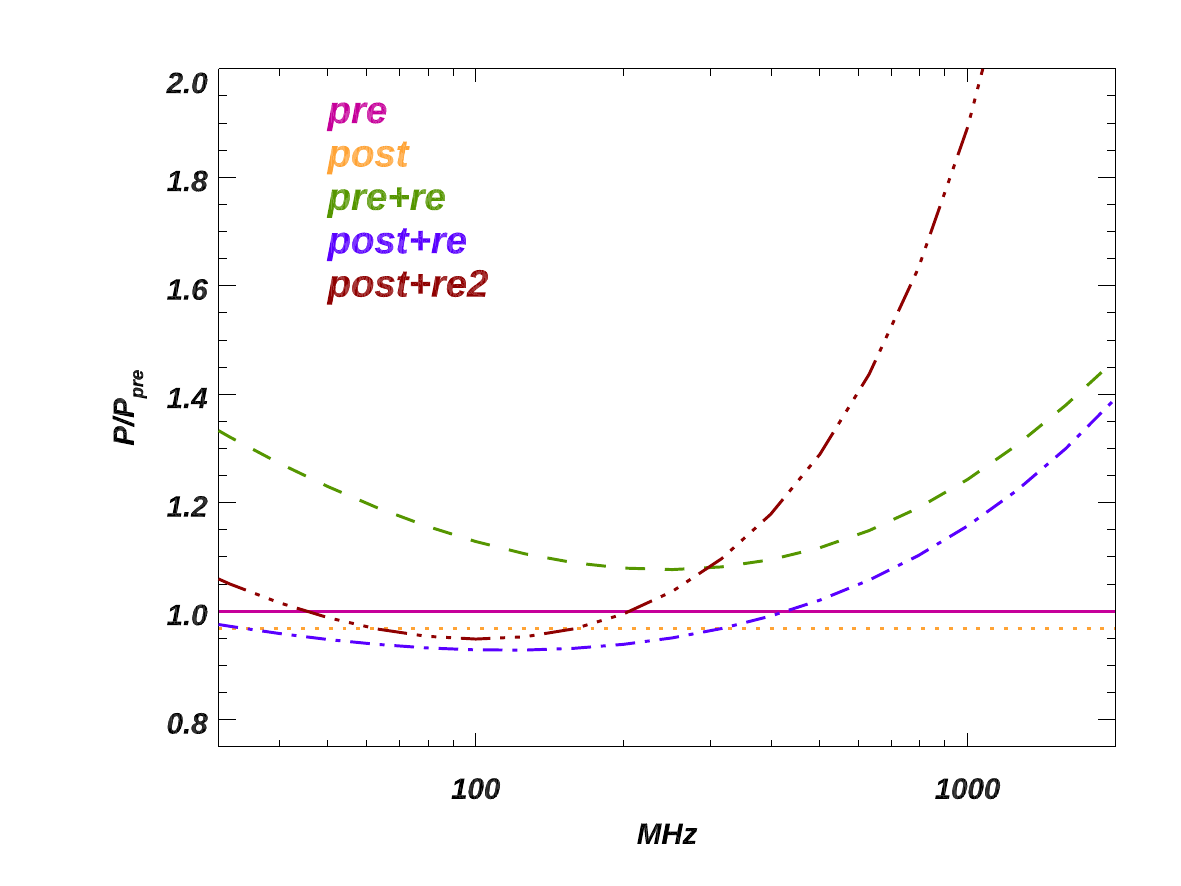}
\caption{Top panel: predicted radio spectra for our models of the radio bridge, as in Tab.\ref{tab:sim}.
Bottom panel: Same as top panel, but with radio emission normalised to the model $pre$ to highlight the differences among the models.}
\label{sim_pred}
\end{figure}   

\begin{table*}

\caption{{ Model parameter for our simulation of the radio bridge (see Sec. \ref{sec:bridge} for more details)}.}
\centering

\tabcolsep 5pt 
\begin{tabular}{c|c|c|c|c|c|c|c|c|c|c|c|c}
\hline
run ID & $n_{\rm post}$ & $T_{\rm post}$ & $\mathcal{M}_{\rm inj}$ & $\log_{\rm 10}(E_{\rm e,inj})$  & $\xi_e,inj$ &$t_{\rm inj}$ & $\mathcal{M}_{\rm re}$ & $t_{\rm re}$ & $B$& $S_{\rm 147 \, MHz}$& $\alpha^{\rm 147 \, MHz }_{\rm 608 \, MHz}$ & $S_{\rm 50 \, MHz}$ \\ 
         &  $cm^{-3}$ & $K$ & & erg & &  Gyr & &  Gyr& $\rm \mu G$  & mJy && mJy\\ \hline \hline

 pre & $1.2 \cdot 10^{-3}$ & $1.6 \cdot 10^{8}$ & 5.0 &  53.45 & $10^{-3}$ & -0.06 & $-$ & $-$ & 3.0 & 13.0 & -1.43 & 49.4 \\
 pre+re & $9 \cdot 10^{-4}$ & $5.3 \cdot 10^{7}$ & 15.0 &  53.90 & $10^{-4}$ & -0.9 & 2.5 & -0.02 & 1.0 & 13.0 & -1.51 & 60.0 \\

 post & $3  \cdot 10^{-4}$ & $1.9 \cdot 10^{7}$ &  5.0 &  53.35 & $3 \cdot 10^{-2}$ & -0.06 & $-$ & $-$ & 3.0 & 13.0 & -1.42 & 47.0 \\

 post+re & $3  \cdot 10^{-4}$ & $1.9 \cdot 10^{7}$ & 15.0 &  52.54 & $1.6 \cdot 10^{-3}$ & -0.9 & 3.0 & -0.04 & 3.0 & 13.0 & -1.42 & 46.0 \\   \hline    
post+re2 & $3 \cdot 10^{-4}$ & $1.9 \cdot 10^{7}$ & 25.0 &  53.59 & $1.6 \cdot 10^{-2}$ & -1.0 & 2.2 & -0.05 & 3.0 & 13.0 & -1.40 & 48.7 \\ 
\hline
\hline

 \end{tabular}
 \label{tab:sim}
\end{table*}

}

\subsection{Radio halo and X-ray emission}
\label{sec:radioX_halo}
Theoretical models for the formation of radio halos would expect that radio emission approximately follows the X-ray emission from the gas \citep[][and ref. therein]{BJ14}.
This is observed in some clusters \citep[e.g.][]{Govoni01} while it is not true in
other  cases (e.g. Abell 1132, \citealt{Wilber17}).
The radio halo in MACSJ0717 is probably the most  striking case where radio emission is offset from the X-ray emission. 
In the turbulent re-acceleration scenario, this would require a different energy in turbulence and magnetic field in regions with and without radio emission, and/or the presence of a seed population of electrons only in the former region.\par
While the latter hypothesis is hard to verify, the amount of energy in turbulence in different cluster regions can be estimated through the amplitude of gas density fluctuations measured from X-ray observations. In stratified cluster atmospheres, the amplitude of gas density fluctuations, $\frac{\delta\rho_k}{\rho}$, and one-component velocity, $V_{1k}$, are proportional to each other at each wavenumber $k$ within the inertial range of scales, namely:
\begin{equation}
\frac{\delta\rho_k}{\rho}=\eta\frac{V_{1k}}{c_s},
\label{eq:ps}
\end{equation}
where $c_s$ is the sound speed of the gas and $\eta=1.0\pm0.3$ is the proportionality coefficient calculated from cosmological simulations of galaxy clusters (\citealt{Zhuravleva14}, see also \citealt{Zhuravleva15} for applications). A similar method has been recently used by \citet{Eckert17}.
Using this approach, we compare the amplitude of the density fluctuations in the regions of the cluster with and without radio halo emission and derive information on the spectrum of the velocity field in the two regions.
We have processed the {\it Chandra} data published by \citet{vanWeeren17}, and  analysed the cluster image in 0.5-3.5 keV band. This band is chosen because the X-ray surface brightness is almost independent on the gas temperature. In order to remove a first order global density gradient, we have fitted the radial profile of the X-ray surface brightness with a spherically symmetric $\beta$ model, and divided the image by this model. We have computed the power spectrum of the X-ray surface brightness fluctuations using a modified $\Delta-$variance method  \citep{Churazov12}. Following  \citet{Churazov12}, the X-ray surface brightness fluctuations have been analysed using a 2D power spectrum approach, and the resulting spectrum in 2D has been then converted to a 3D power spectrum of gas density fluctuations.
{ The cluster X-ray emission is complex, and a spherically symmetric $\beta$-model is not an accurate description of the gas distribution. Nonetheless, it permits to remove a first-order density gradient. The results that follow depend on the underlying model. In Sec. \ref{sec:dep}, we discuss how our assumption affects the results.}

\begin{figure}
\vspace{100pt}
\begin{picture}(150,150)
\put(0,0){\includegraphics[width=\columnwidth]{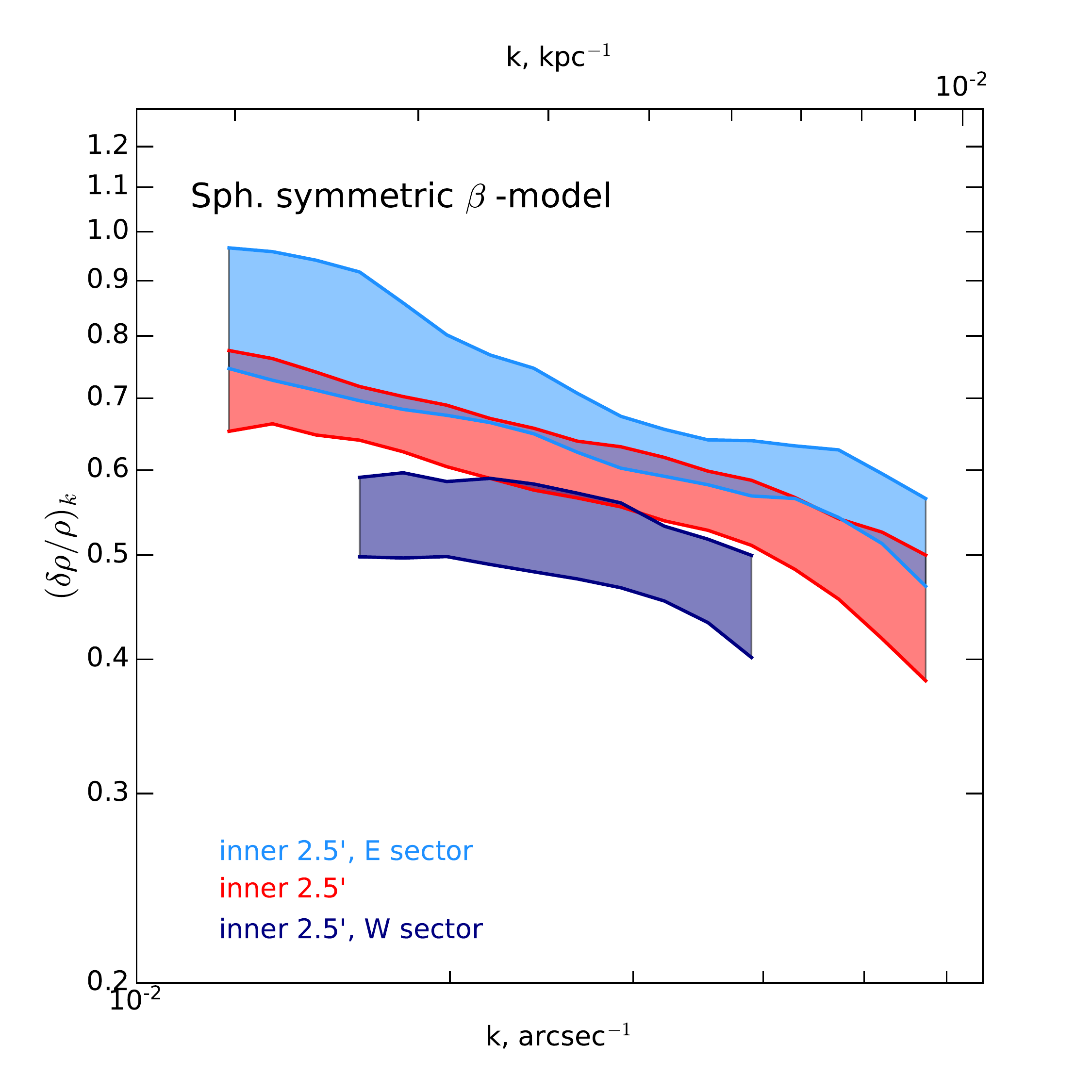}}
\put(151,27){\includegraphics[width=2.2cm]{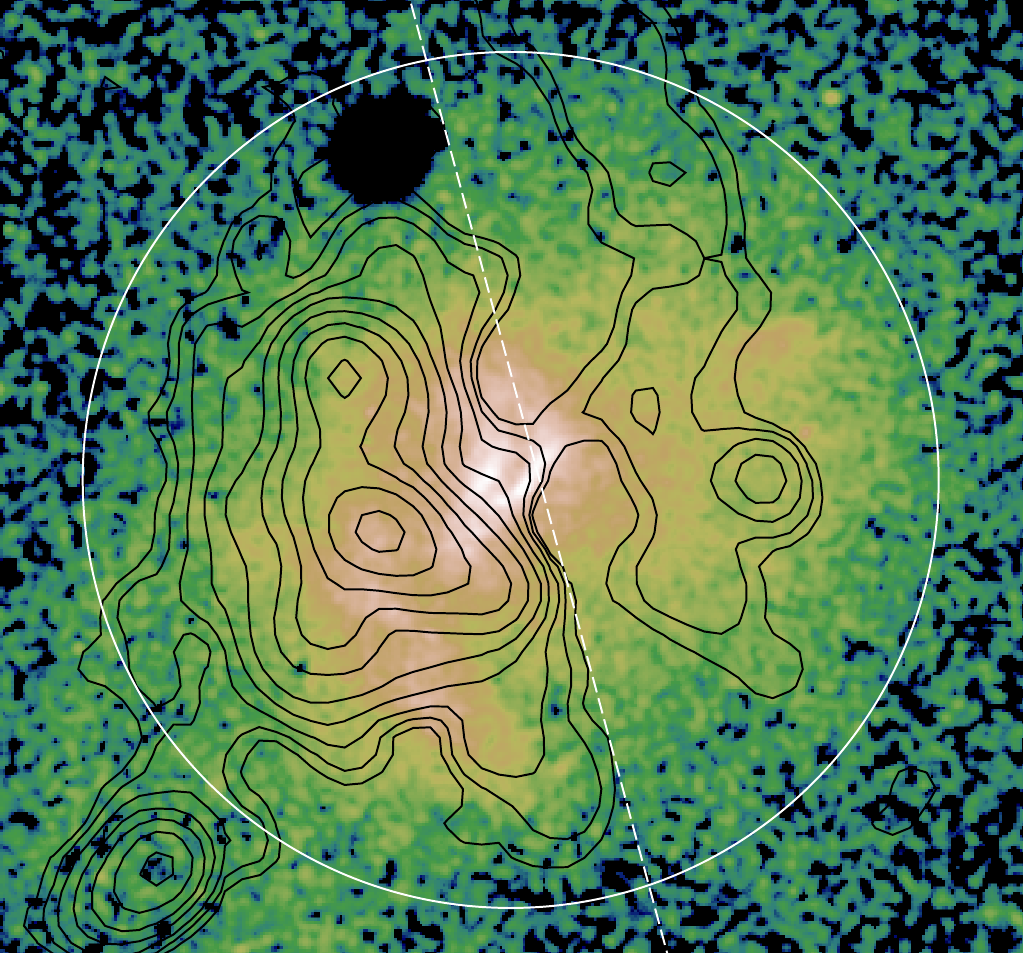}}
\end{picture}
\caption{
Red region: amplitude of the gas density fluctuations  as a function of the wavenumber $k$ derived for the whole cluster emission (white circle in the inset). Purple and light blue regions: amplitude of the gas density fluctuations 
computed for the E and W part, respectively. }
\label{fig:PS}
\end{figure}

 In Fig. \ref{fig:PS}, we show the amplitude of density fluctuations  as a function of wavenumber $k$ in the two regions of the clusters.
On a scale of $\sim$ 350 kpc, the average amplitude of density fluctuations is $0.77 \pm
0.09$ in the region of the halo and  $0.55 \pm 0.05$ in the region without radio halo. 
This gives a ratio of $\sim$ 1.4 between the two. If eq. \ref{eq:ps} holds, neglecting the differences in density, and assuming that $c_s$ is the same in the two regions, we can conclude that the ratio of kinetic over thermal energy is twice as large in the region with radio emission than in the region without. However, there are indications that the temperature in the two regions is different by a factor $\sim 1.4$
\citep{vanWeeren17}, which would translate in a factor 3 of ratio of kinetic over thermal energy in the regions with and without radio emission.
Interestingly, the ratio of the average radio power at 147 MHz in the regions with and without radio emission is more than a factor 40 (at 1$\sigma$). 
{ Neglecting effects due to different magnetic fields and/or populations of seed particles in the two regions, our results suggest that the power emitted by electrons in the radio halo  at 147 MHz has a super-linear scaling with the gas kinetic energy.  }

{
\subsection{Underlying model of the X-ray surface brightness.}
\label{sec:dep}
The results obtained in the previous section depend on the assumptions we have done on the X-ray surface brightness distribution of the cluster.
We have modelled the X-ray surface brightness using a spherically symmetric $\beta$-model, which is a good representation for virialised systems. As the cluster is
in a very active merger state, non-negligible departures from a spherically symmetric $\beta$-model are expected. This can be seen from the X-ray image shown in Fig. 
\ref{fig:Xradio}, and also in Fig. \ref{fig:beta_residuals}, where we show the residuals of the X-ray surface brightness after division by the spherically symmetric $\beta$-model. 

To check how the asymmetry of the gas distribution affects the density amplitude measurements, we repeated the analysis of Sec. \ref{sec:radioX_halo} considering a different $\beta$-model (so called ``patched" $\beta$-model, \citealt{Zhuravleva15}), which is elongated in the SE-NW direction. Our patched $\beta$-model is defined as in \citet{Zhuravleva15}, i.e. $I_{pm}= I_{\beta}S_{\sigma} [I_X/I_{\beta}]$, where $I_{\beta}$ is the spherically symmetric $\beta$-model, $I_X$ is the cluster X-ray surface brightness, $S_{sigma}[  \cdot ]$ is the Gaussian
smoothing with the smoothing window size $\sigma$. We choose $\sigma=50$, and the resulting patched $\beta$-model is shown in Fig. \ref{fig:beta_residuals} (top right panel). In the same figure, we also plot the residuals of the X-ray surface brightness distribution, obtained dividing the X-ray image by the patched $\beta$-model (bottom right panel).
In Fig \ref{fig:PS_mod}, we show the amplitude of the gas density fluctuations as a function of the wavenumber $k$, obtained assuming a patched $\beta$-model instead of a spherically symmetric $\beta$-model.  The amplitude of the  gas density fluctuations is suppressed in both the E and W regions, indicating a strong dependence on the underlying model.

On a scale of $\sim$ 350 kpc, the average amplitude of density fluctuations is now  0.47$\pm$0.02 in the region of the halo and  0.41$\pm$0.04 in the region without radio halo. Hence, they are consistent within 1 $\sigma$.
If we only consider the mean value, we obtain a ratio of  $\sim$ 1.13  for the two regions, hence  $v_{kin}$ that is 30$\%$ higher in the region of the radio halo.

\par
This analysis indicates that our results depend on the model of gas density that we use.
Either using a spherically symmetric and a patched $\beta$-model, the amplitude of the gas density fluctuations are higher in the E region than in the W region, although when we consider
the patched $\beta$-model the difference is only marginal, and the amplitudes are consistent within 1$\sigma$.

Constraining the dependence of the radio power of the gas kinetic energy would give important constraints for theoretical models of halo formation (e.g. \citealt{BrunettiLazarian16}). The analysis we have performed
here suggests a super-linear scaling of the radio power with respect to the gas kinetic energy, as well as a strong dependence of the results on the underlying model for the cluster gas density distribution.

\begin{figure}
\includegraphics[width=\columnwidth]{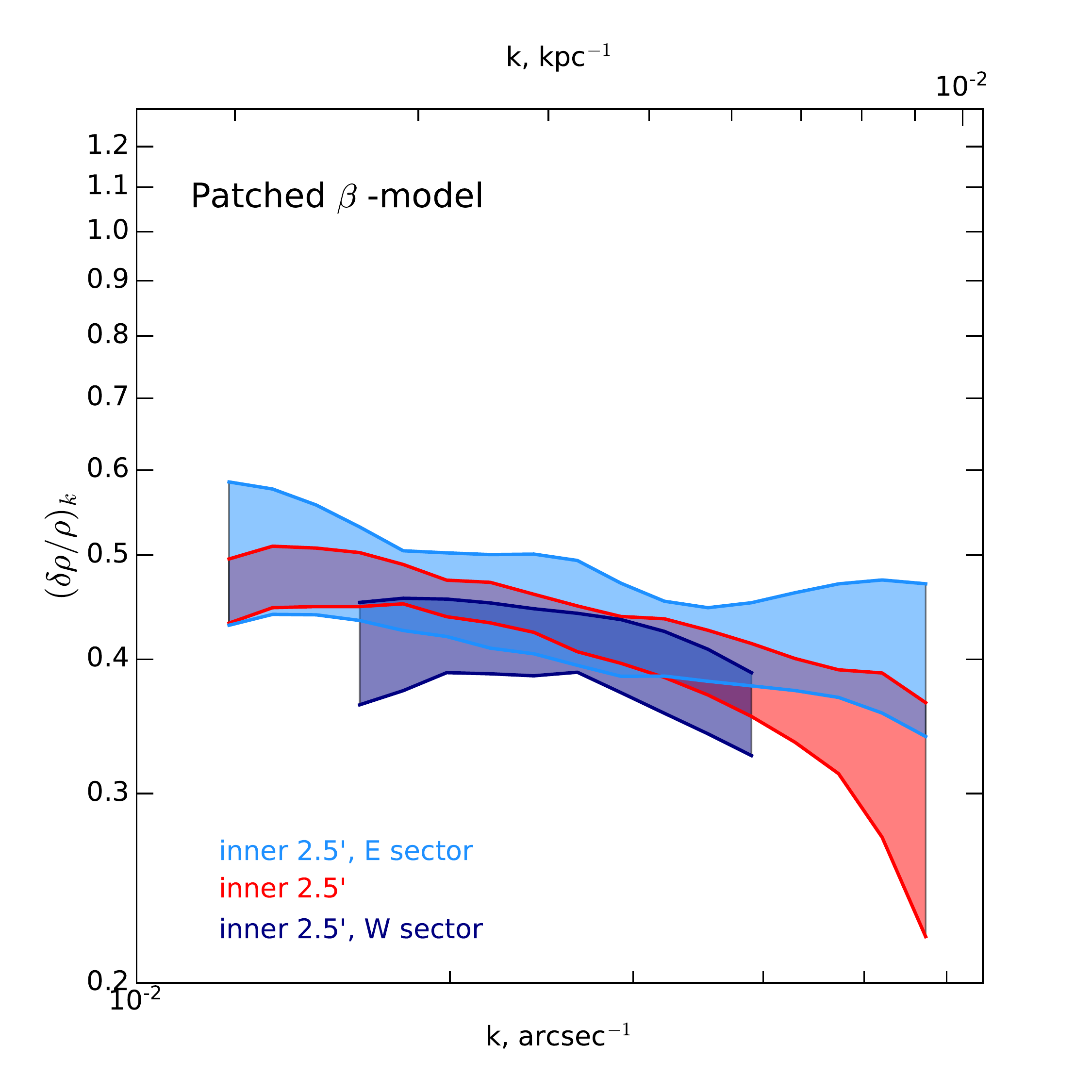}

\caption{Same as Fig. \ref{fig:PS} but considering a patched $\beta$-model for the cluster density distribution. Red region: amplitude of the gas density fluctuations  as a function of the wavenumber $k$ derived for the whole cluster emission. Purple and light blue regions: amplitude of the gas density fluctuations computed for the E and W part, respectively. }
\label{fig:PS_mod}
\end{figure}

\begin{figure}

\includegraphics[width=\columnwidth]{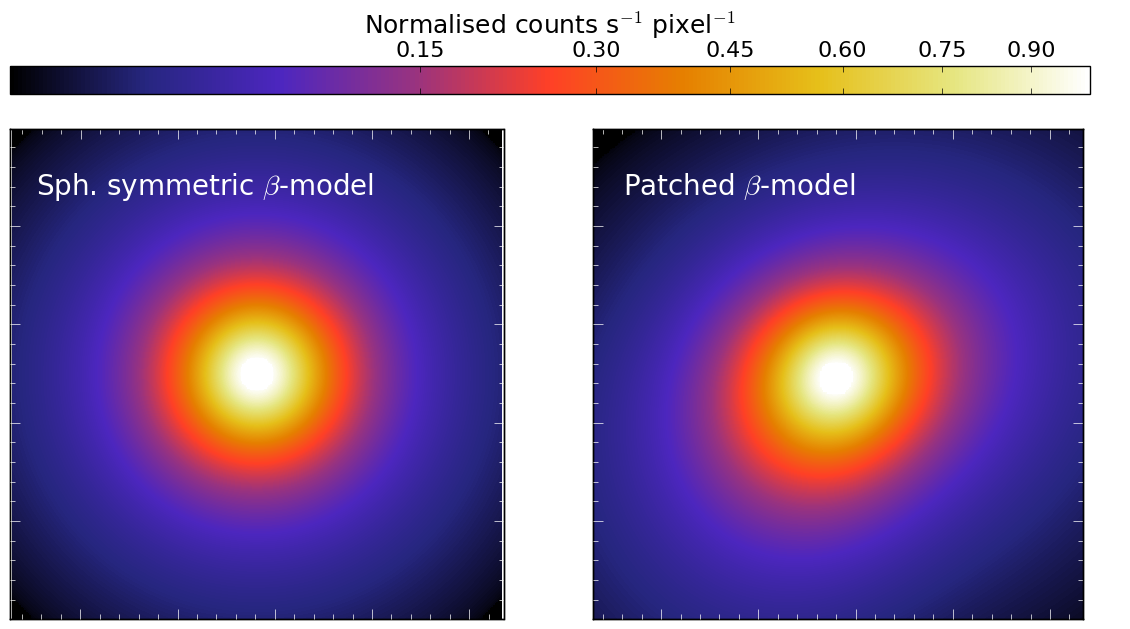}
\includegraphics[width=\columnwidth]{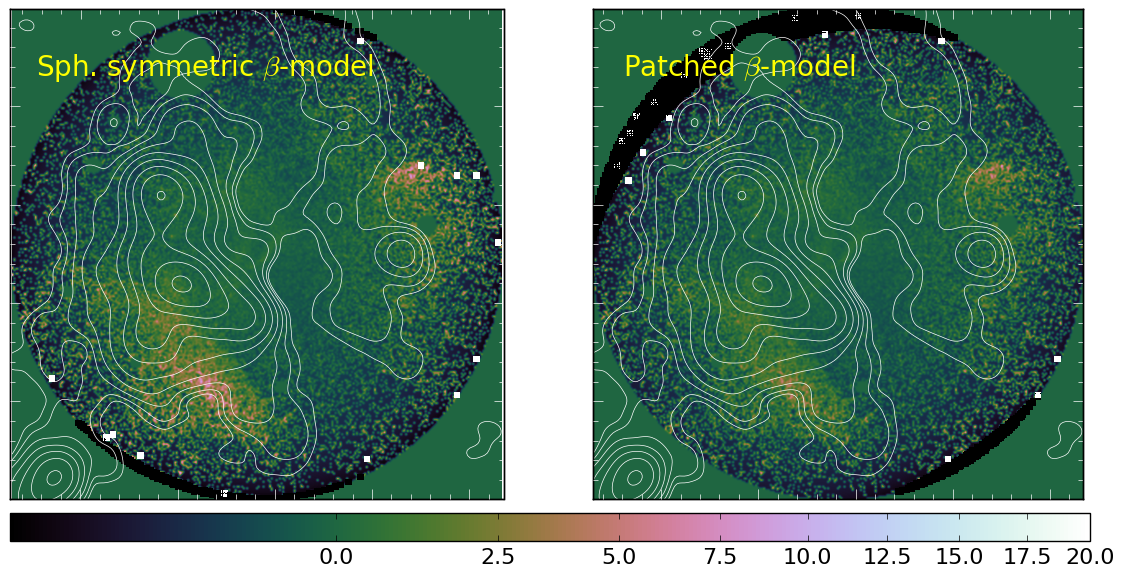}
\caption{Top panels: Normalised spherically symmetric $\beta$-model (left) and patched $\beta$-model (right) used to derive a first-order density gradient of the gas.
Bottom panels: residuals of the X-ray emission after division by the spherically symmetric $\beta$-model (left) and patched $\beta$-model (right).}
\label{fig:beta_residuals}
\end{figure}

}

\section{Conclusions}
\label{sec:discussion}
MACSJ0717 is undergoing a violent merger that involves at least four sub-clusters. The radio emission is complex and shows unique features that are visible from 147 MHz up to 5 GHz.
We have presented new results from LOFAR and GMRT observations of the galaxy cluster MACSJ0717, and using X-ray observations, we have derived new constraints on the particle acceleration processes
in the cluster centre and outskirts. 
Our results can be summarised as follows:
\begin{itemize}
\item{LOFAR observations at 147 MHz reveal new emission from the ICM: (i) a radio arc located NW of the cluster centre and extending for 1.9 Mpc in the NS direction; (ii) a radio bridge connecting the main cluster to a HT radiogalaxy located in the direction of a 19 Mpc-long filament of galaxies; (iii) a radio bar, that traces the X-ray bar observed S of the main mass component. }
\item{Using GMRT observations at 608 MHz, we have constrained the spectra of the radio arc ($\alpha^{\rm 147 \, MHz }_{\rm 608 \, MHz} < -1.3$), the radio bridge ($\alpha^{\rm 147 \, MHz }_{\rm 608 \, MHz} < -1.4$), and the radio bar ($\alpha^{\rm 147 \, MHz }_{\rm 608 \, MHz} < -1.4$). The spectra of the radio halo and of the radio relic do not show significative departure from the power-law observed at higher frequencies and already studied in the literature.}
\item{ We have investigated under which conditions the radio bridge can originate from electron re-acceleration by a weak Mach number shock.
A ``transversal" shock moving perpendicular to the filament main axis can explain the properties of the radio emission, although some fine tuning of the parameters is required.}
\item{The radio halo at LOFAR frequencies is more extended than previously observed. Nonetheless, the radio emission does not follow the X-ray emission of the gas in the W part of the cluster. 
Assuming that the spectrum of density fluctuations -- as deduced from {\it Chandra} observations -- traces the spectrum of the gas velocity, data suggest a different ratio of kinetic over thermal energy in the regions with and without radio halo. { This result depend on the model we assume for the gas density distribution, and lacks of a robust statistical significance. Deeper observations, as well as a more accurate modeling for the cluster density distribution would be required to investigate this point in more detail. }}

\end{itemize}

\section*{Acknowledgments}
Annalisa Bonafede acknowledges financial support from the ERC-Stg DRANOEL, no 714245.
FV acknowledges financial support from the ERC-Stg MAGCOW, no.714196. 
RJvW and HJAR acknowledge support from the ERC Advanced Investigator programme NewClusters 321271 
and the VIDI research programme with project number 639.042.729, which is financed by the Netherlands Organisation for Scientific Research (NWO).
We thank Prof. Daniele Dallacasa for useful discussions, and the referee for useful comments.
This work had made use of the Lofar Solution Tool (LoSoTo), developed by F. de Gasperin.
F.d.G. is supported by the VENI research programme with project number 1808, which is financed by the Netherlands Organisation for Scientific Research (NWO)
We thank the staff of the GMRT that made these observation possible. GMRT is run by the National Centre for
Astrophysics of the Tata Institute of Fundamental Research. 

This research has made use of the NASA/IPAC Extragalactic Database (NED), which is operated by the Jet Propulsion Laboratory, California Institute of Technology, under contract with the National Aeronautics and Space Administration.
\bibliographystyle{mn2e}
\bibliography{master}

\label{lastpage}

\end{document}